\documentclass[letterpaper,twocolumn,10pt]{article}
\usepackage{preprint}
\usepackage{authblk}
\usepackage[
  margin=0.5in,
  bottom=0.875in,
  columnsep=0.25in,
]{geometry}
\usepackage{microtype}
\usepackage{stfloats}

\usepackage[T1]{fontenc}
\usepackage{helvet}

\usepackage{amsmath}
\usepackage{amssymb}
\usepackage{mathtools}
\usepackage{nicefrac}
\usepackage{optidef}
\usepackage[
  separate-uncertainty=true,
  multi-part-units=single,
  list-units=repeat,
  detect-weight,
  mode=text
]{siunitx}

\usepackage{booktabs}
\usepackage{multirow}
\usepackage{graphicx}
\usepackage{enumitem}
\usepackage[normalem]{ulem}

\usepackage[tiny,compact]{titlesec}
\titleformat{\abstract}[runin]{\bfseries}{\theabstract}{1em}{}
\titleformat{\section}{\bfseries}{\thesection}{1em}{}
\titleformat{\subsection}[runin]{\bfseries}{\thesubsection}{1em}{}
\titleformat{\paragraph}[runin]{\bfseries}{\thesubsection}{1em}{}

\usepackage[font={small}]{caption}
\usepackage[numbers,sort,compress]{natbib}
\usepackage[sectionbib]{bibunits}
\defaultbibliographystyle{unsrtnat}
\defaultbibliography{references}

\usepackage[usenames,dvipsnames,table,xcdraw]{xcolor}
\usepackage[bookmarks,colorlinks,breaklinks]{hyperref}
\hypersetup{linkcolor=red, citecolor=cyan}
\usepackage{xurl}

\usepackage[capitalize]{cleveref}
\crefname{section}{Sec.}{Secs.}
\Crefname{section}{Section}{Sections}
\Crefname{table}{Table}{Tables}
\crefname{table}{Tab.}{Tabs.}
\Crefname{meth}{Methods}{Methods}

\usepackage{xspace}
\makeatletter
\DeclareRobustCommand\onedot{\futurelet\@let@token\@onedot}
\def\@onedot{\ifx\@let@token.,\else.,\null\fi\xspace}
\DeclareRobustCommand\justonedot{\futurelet\@let@token\@justonedot}
\def\@justonedot{\ifx\@let@token.\else.\null\fi\xspace}
\def\eg{\emph{e.g}\onedot} 
\def\ie{\emph{i.e}\onedot} 
 
\def\etc{\emph{etc}\justonedot} \def\vs{\emph{vs}\justonedot}
\makeatother

\DeclareMathOperator{\pKi}{\ensuremath{\prescript{P}{}{\mathbf K}_{\mathit{I}}}}
\DeclareMathOperator{\iKc}{\ensuremath{\prescript{I}{}{\mathbf K}_{\mathit{C}}}}
\DeclareMathOperator{\cTw}{\ensuremath{\prescript{C}{}{\mathbf T}_{\mathit{W}}}}
\DeclareMathOperator{\wTc}{\ensuremath{\prescript{W}{}{\mathbf T}_{\mathit{C}}}}
\DeclareMathOperator{\SO3}{\ensuremath{\mathbf{SO}(3)}}
\DeclareMathOperator{\SE3}{\ensuremath{\mathbf{SE}(3)}}
\DeclareMathOperator{\so3}{\ensuremath{\mathfrak{so}(3)}}

\newcommand{\name}{\textbf{\textit{xvr}}\xspace}
\newcommand{\qtyci}[5]{\qty{#1}{#2} (#3: \num{#4}--\num{#5})}

\definecolor{myblue}{HTML}{47a3ff}
\definecolor{myorange}{HTML}{ff6600}
\definecolor{mypink}{HTML}{ff66b3}

\newcommand{\blue}{{\color{myblue}\textbf{blue}}\xspace}

\newcommand{\pink}{{\color{mypink}\textbf{pink}}\xspace}

\title{Rapid patient-specific neural networks for intraoperative X-ray to volume registration}
\author[1,2,3]{Vivek Gopalakrishnan$^\dagger$}
\author[3]{David-Dimitris Chlorogiannis}
\author[4]{Andrew Abumoussa}
\author[5]{Anna M. Larson}
\author[3]{Nazim Haouchine}
\author[6]{Darren B. Orbach}
\author[3]{Sarah Frisken}
\author[2,3,7]{Neel Dey$^\dagger$}
\author[1,2]{Polina Golland$^\dagger$}
\affil[1]{Harvard-MIT Health Sciences and Technology, Massachusetts Institute of Technology, Cambridge, MA, USA}
\affil[2]{Computer Science and Artificial Intelligence Laboratory, Massachusetts Institute of Technology, Cambridge, MA, USA}
\affil[3]{Department of Radiology, Harvard Medical School, Boston, MA, USA}
\affil[4]{Saint Luke's Marion Bloch Neuroscience Institute, Kansas City, MO, USA}
\affil[5]{Department of Critical Care Medicine, Shriners Children's Hospital, Boston, MA, USA}
\affil[6]{Department of Interventional Neuroradiology, Boston Children's Hospital, Boston, MA, USA}
\affil[7]{Athinoula A. Martinos Center for Biomedical Imaging, Massachusetts General Hospital, Boston, MA, USA}
\affil[$\dagger$]{\it Correspondence to: \href{mailto:vivekg@csail.mit.edu,dey@csail.mit.edu,polina@csail.mit.edu}{\{vivekg,dey,polina\}@csail.mit.edu}}

\abstract{Advanced navigation techniques in image-guided interventions and surgical robotics require the rapid and precise alignment of 3D preoperative volumes (\eg CT, MRI) to 2D intraoperative images (\eg X-ray fluoroscopy)~\cite{unberath2021impact, yip2023artificial}. However, existing 2D/3D registration methods fail to generalize across the broad spectrum of fluoroscopy-guided procedures: traditional intensity-based optimizers require careful hyperparameter tuning for each subject~\cite{penney1998comparison, knaan2003effective}, while deep learning approaches demand extensive manually labeled datasets and remain constrained to the specific anatomy on which they were trained~\cite{grupp2020automatic, grimm2021pose}. To address these limitations, we present \name, a self-supervised framework that combines patient-specific neural networks with gradient-based optimization for automatic 2D/3D registration. \name leverages physics-based simulation to generate training data from a patient's own preoperative scan, eliminating the need for manual annotation. We present a foundation model pretrained on thousands of whole-body scans, achieving patient-specific adaptation for any anatomical region in only \qty{5}{\minute} of finetuning. In the largest evaluation of 2D/3D registration on real fluoroscopy to date, \name achieves high accuracy in seconds across diverse anatomical structures, imaging modalities, and hospitals, improving upon the accuracy of existing methods by an order of magnitude. \name makes pan-anatomical 2D/3D rigid registration accessible to broad clinical and research communities through open-source software at \url{https://xvr.csail.mit.edu}.}

\begin{document}
\begin{bibunit}
\maketitleandabstract

\begin{figure*}[hbt!]
    \centering
    \includegraphics[width=\linewidth]{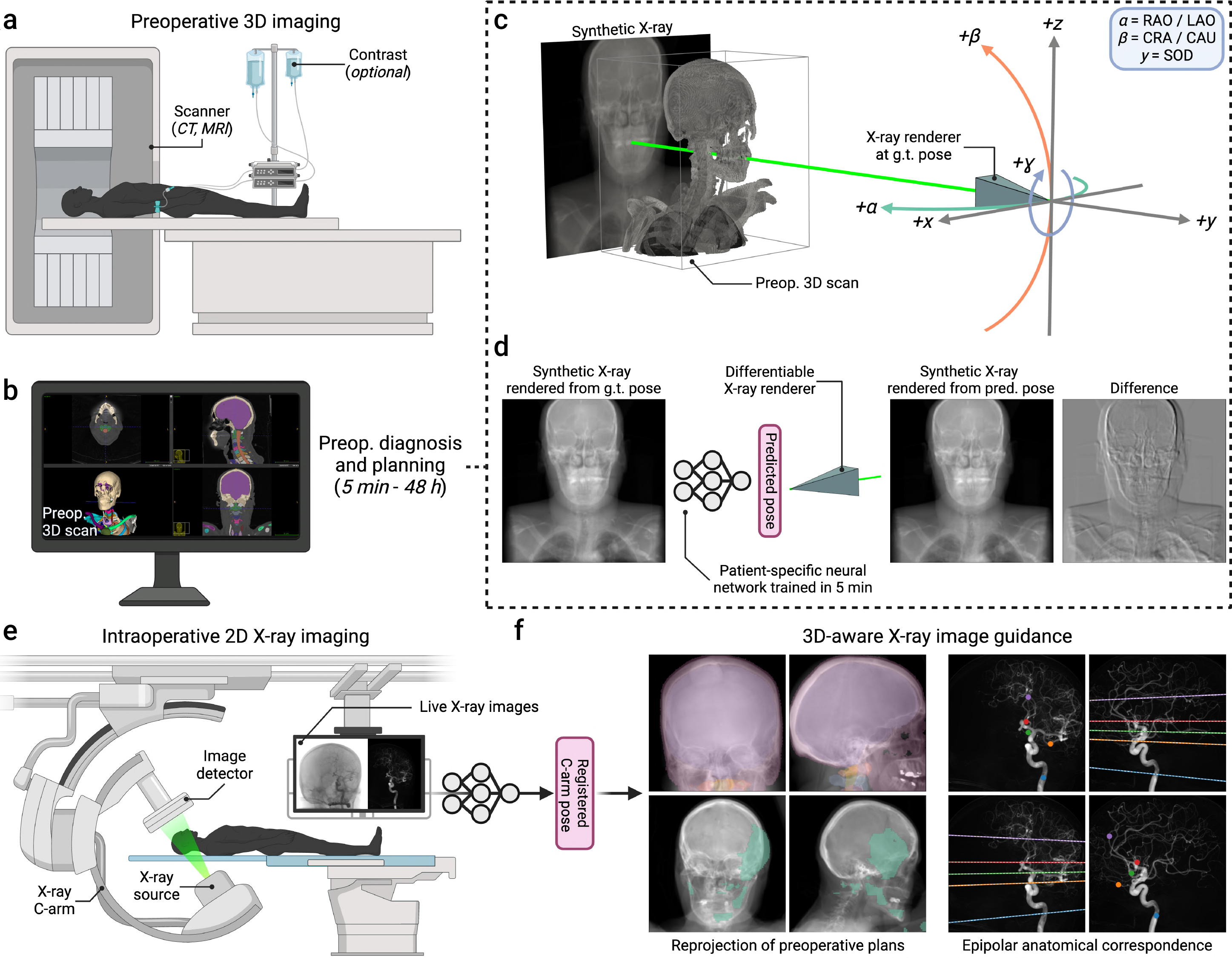}
    \caption{\textbf{\name rapidly trains patient-specific neural networks that achieve accurate intraoperative 2D/3D registration while integrating into existing clinical workflows.} \textbf{a,} Preoperative 3D imaging is routinely acquired before image-guided procedures. \textbf{b,} Clinical teams make diagnoses and preoperative plans from these scans, with planning times ranging from minutes to days depending on the intervention (\eg stroke \vs radiotherapy). \textbf{c,}~Preoperatively, we render synthetic X-rays from the patient's 3D imaging using our differentiable renderer, whose geometry conforms with the radiologic nomenclature of commercial C-arms: rotations $\alpha$ and $\beta$ correspond to the left-right anterior oblique (LAO/RAO) and the craniocaudal (CRA/CAU) axes, respectively, and translation $y$ corresponds to the source-to-object distance (SOD). \textbf{d,}~We finetune patient-specific networks to regress the ground truth (g.t.) pose of these synthetic X-rays in just \qty{5}{\minute}. \textbf{e,} Patient-specific networks perform accurate 2D/3D registration on live 2D X-rays to 3D preoperative scans in seconds. \textbf{f,}~This enables numerous applications, including reprojection of 3D preoperative plans or anatomical targets onto intraoperative imaging or identification of shared structures across multiple X-ray views using epipolar geometry.}
    \label{fig:overview}
    \vspace{-1em}
\end{figure*}

\section{Main}
Each year, millions of clinical interventions are performed under real-time X-ray guidance~\cite{mahesh2022patient}. While C-arm imaging devices enable noninvasive visualization of interventions from virtually any angle, X-rays are projection images that do not provide explicit depth information. This spatial ambiguity confounds the three-dimensional (3D) navigation of medical devices within the body, increasing the risk of suboptimal device deployment and intraoperative complications~\cite{cornelis2023imaging, jhawar2007wrong}. Although volumetric modalities such as CT and MRI are routinely acquired preoperatively, they are often unavailable during procedures due to their lengthy acquisition times and incompatibility with surgical workflows~\cite{tonetti2020role}. A promising alternative is to emulate volumetric image guidance by aligning two-dimensional (2D) intraoperative X-rays to 3D preoperative scans, enabling the localization of surgical devices relative to 3D anatomy. This capability makes 2D/3D registration critical to the development of next-generation image navigation and robotic surgery technologies~\cite{abumoussa2023machine, naik2022hybrid, metz2009patient, wagner20164d, huynh2020artificial, kim2022telerobotic}.

Despite its broad potential, achieving reliable 2D/3D X-ray to volume registration across diverse specialties remains a challenge~\cite{unberath2021impact}. Conventional methods estimate a C-arm's pose by combining iterative optimization with computational models of X-ray formation, searching for the synthetic X-ray rendered from the preoperative volume that most closely matches the real intraoperative X-ray~\cite{penney1998comparison, knaan2003effective}. Although accurate, iterative optimization is sensitive to initialization errors: if started even a few centimeters from the true pose, iterative solvers often converge to incorrect solutions~\cite{gu2020extended, gopalakrishnan2022fast, gao2023fully}. To this end, numerous deep learning approaches have been developed to produce more accurate initializations, either by using keypoint estimators to identify paired anatomical landmarks in the 2D and 3D images~\cite{grupp2019pose, bier2019learning, shrestha2024rayemb} or by directly regressing the C-arm pose from an X-ray~\cite{miao2016cnn, bui2017x, zhang2023patient}. While typically more accurate than direct pose regression, supervised keypoint estimation methods require expert knowledge of reliably visible anatomical landmarks, as well as manual annotation of these structures for every new preoperative 3D scan~\cite{grupp2020automatic}. Even semi-supervised approaches that transfer 3D landmarks from previous patients require a large training corpus of manually annotated scans~\cite{grimm2021pose}.

We introduce \name, a self-supervised framework for patient-specific 2D/3D rigid registration that generalizes to any patient, procedure, or pathology~(\cref{fig:overview}). Preoperatively, a network is trained to regress the poses of synthetic 2D X-rays rendered from randomly simulated virtual C-arms from the patient's own preoperative scan. Intraoperatively, this network provides an accurate initial C-arm pose that can be rapidly refined by a gradient-based optimizer via differentiable X-ray rendering. Unlike previous (semi-)supervised methods, \name leverages patient-specific simulation to generate unlimited synthetic X-rays with ground truth C-arm poses. To enable rapid deployment across diverse clinical scenarios, we pretrain a foundation model with synthetic X-rays generated from over 2,000 volumes with whole-body and \ang{360} C-arm pose coverage. Finetuning this model on a new patient reduces training time from hours~\cite{gopalakrishnan2024intraoperative} to only \qty{5}{\minute}.

Analysis across five datasets from different hospitals spanning multiple patient populations, anatomical structures, and interventional specialties establish three key findings. First, \name outperforms previous (semi-)supervised deep learning and unsupervised iterative methods in both accuracy and robustness. Second, patient-specific finetuning eliminates the out-of-distribution failures that previous methods suffer when encountering patients not well-represented in their training sets, ensuring reliable performance for every individual. Third, \name operates at intraoperative speeds with consistent accuracy across heterogeneous real-world clinical and surgical procedures.

\begin{figure*}[hbt!]
    \centering
    \includegraphics[width=\linewidth]{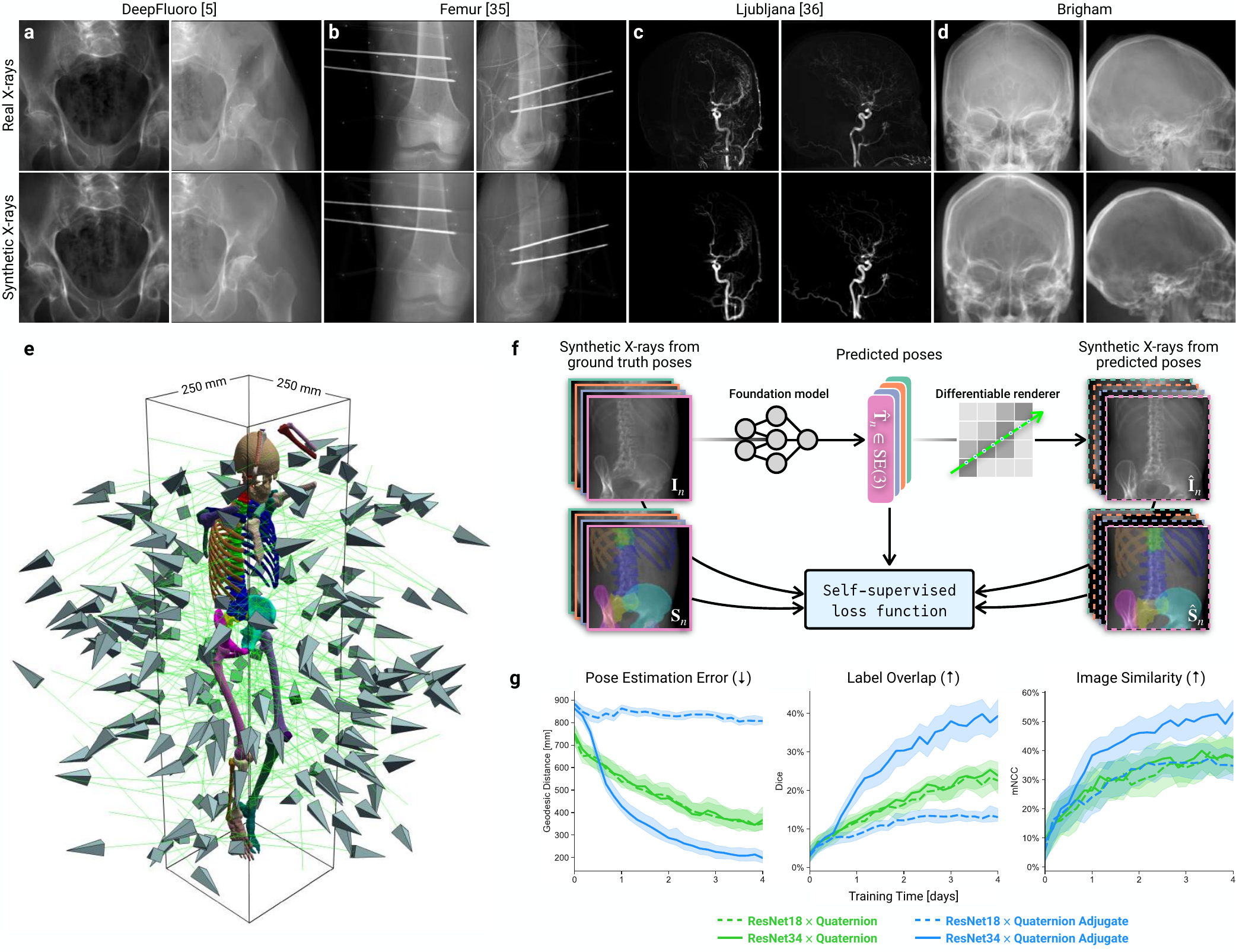}
    \caption{\textbf{Training a foundation model for 2D/3D registration using physics-based differentiable rendering}. \textbf{a-d,} Comparisons of real X-rays to synthetic images rendered from CT scans of the same patients using ground truth C-arm poses demonstrate the fidelity of \name. \textbf{e,} Given a volume, we render synthetic X-rays $\mathbf I_n$ and 2D segmentation masks $\mathbf S_n$ from randomly sampled poses $\mathbf{T}_n$. \textbf{f,} Given an input synthetic X-ray, a pose regression network predicts pose $\mathbf{\hat T}_n$, from which we re-render predicted images $\mathbf{\hat I}_n$ and segmentations $\hat{\mathbf S}_n$. A self-supervised pose regression loss compares all predicted quantities to their ground truth counterparts, which are known as they are simulated. \textbf{g,} Optimizing the pose regression loss requires a sufficiently parameterized network (\ie ResNet34) and a continuous representation of $\SE3$ (\ie quaternion adjugate). Error bars represent 90\% confidence intervals of held-out synthetic data.}
    \label{fig:wbct}
    \vspace{-1em}
\end{figure*}

\section{Differentiable synthetic data synthesis}
To generate synthetic training data from a patient's own preoperative imaging, we developed a differentiable model of X-ray image formation. Given a preoperative 3D volume, \name uses differentiable implementations of ray casting algorithms to render synthetic X-rays from input C-arm poses (\cref{sec:diffdrr}). Optionally, given a 3D label map of the input scan, \name casts samples from voxels with different labels to separate 2D image channels, enabling the rendering and registration of individual anatomical structures. \name's outputs are geometrically consistent with real X-ray images~(\cref{fig:wbct}a-d) with differences in appearance primarily driven by motion between pre- and intraoperative imaging (\eg the left femur moves in the second column of \cref{fig:wbct}a). Nevertheless, the 3D pose of any individual rigid body is identifiable and can be used as the registration target. Finally, \name is highly optimized, capable of rendering thousands of synthetic X-rays per minute, enabling both scalable foundation model pretraining and rapid patient-specific adaptation.

\section{A foundation model for pose regression}
To train a model that generalizes to X-rays of any anatomical region, we aggregated a corpus of 2,245 CT and MR volumes from four public datasets~\cite{wasserthal2023totalsegmentator, jaus2024towards, nitrc2017mra, liu2021deep} that collectively cover the whole body. The scans range broadly from research studies to acute hospitalizations and contain diverse clinical findings, including fractures, metal implants, and contrast enhancement. As these scans were initially not in a common coordinate frame, we rigidly aligned~\cite{yushkevich2016ic} them to an arbitrary template whole-body CT scan~(\cref{fig:wbct}e), enabling the training of a foundation model for pose regression.

Our self-supervised training loop is illustrated in~\cref{fig:wbct}f. In each iteration, we randomly select a scan from the pretraining corpus and sample a batch of C-arm poses, from which we differentiably render synthetic X-rays and their corresponding 2D segmentations. To adapt to any imaging setup, poses are sampled in \ang{360} about the LAO/RAO axis and $\pm$\ang{60} about the CRA/CAU axis~(\cref{fig:overview}c), along with large ranges for the in-plane rotation and three translations (\cref{tab:parameters}). We train a convolutional neural network (CNN) to predict the pose of each input X-ray, from which we re-render predicted X-rays and segmentations. Since all training data is simulated, we evaluate predicted quantities against their corresponding ground truth~(\cref{sec:simulation}). Patient-specific training follows an identical procedure, but uses only the patient's preoperative scan instead of the pretraining corpus.

To train a model that fits this large pose distribution, we identified two critical design choices~(\cref{fig:wbct}g). First, since the space of rigid transforms forms a manifold, $\SE3$, direct regression is impossible with standard CNN architectures, requiring regression onto a Euclidean representation of 3D rotations. While Euler angles and quaternions are common choices, both contain discontinuities that hinder rotation estimation with deep learning~\cite{zhou2019continuity, geist2023rotations}. Instead, we regress the quaternion adjugate, a higher-order representation that resolves these discontinuities~\cite{lin2023algebraically}. Second, sufficient network capacity is essential: a ResNet18 backbone fails to fit the training data, whereas a ResNet34 succeeds. Notably, neither backbone achieves sufficient accuracy when trained to regress quaternions directly. Foundation models were trained for 300,000 steps on a single NVIDIA H200 GPU. Computational resources are detailed in \cref{sec:compute}.

\begin{figure*}[htbp!]
    \centering
    \includegraphics[width=\linewidth]{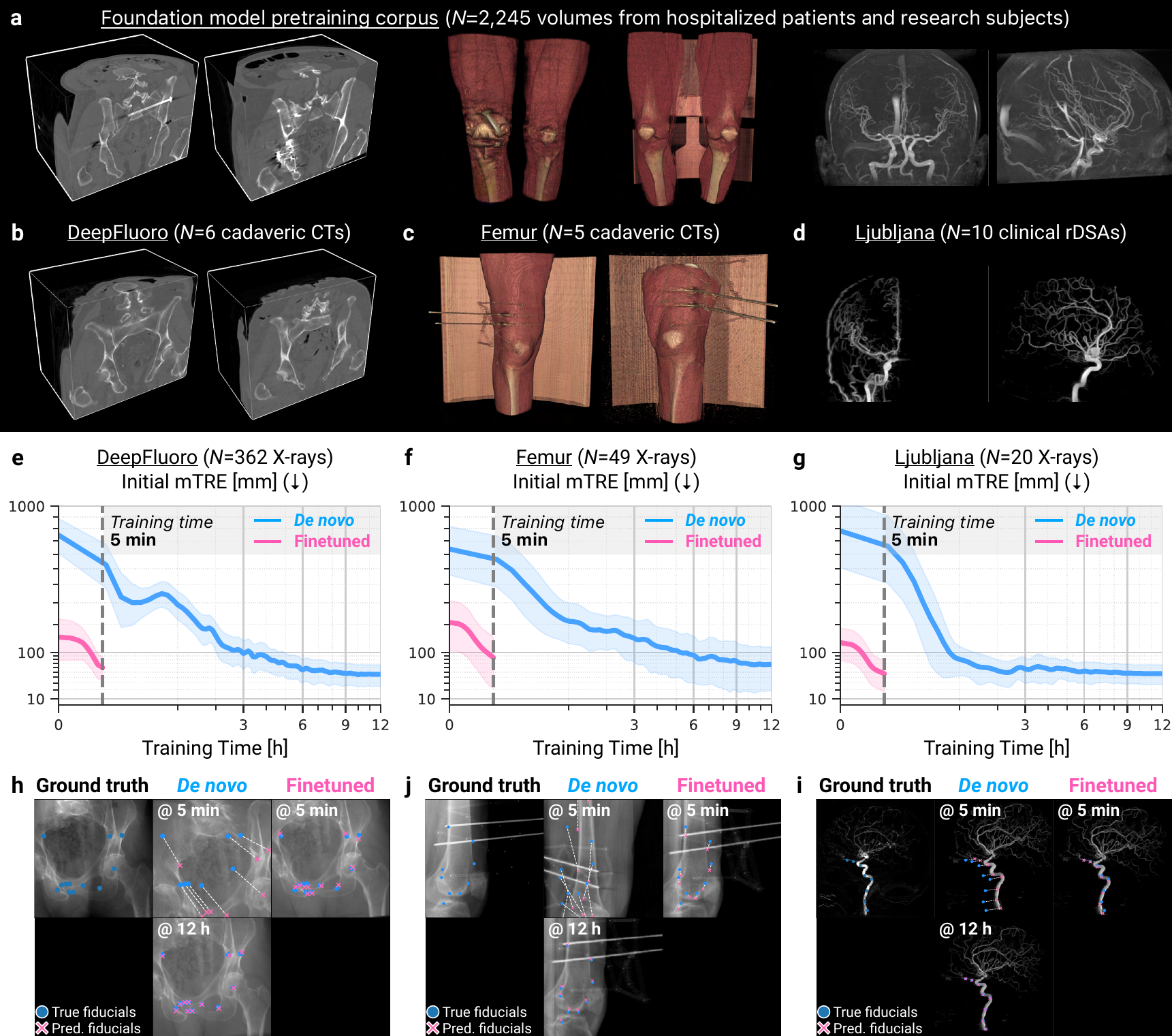}
    \caption{\textbf{Our foundation model for 2D/3D registration enables patient-specific finetuning in \qty{5}{\minute}.} \textbf{a,} Volumes from our foundation model's whole-body pretraining corpus, cropped to match anatomical regions in our benchmark datasets. Our pretraining corpus comprises multiple modalities and diverse clinical findings including fractures, metal implants, and varied acquisition protocols. This sample size and diversity enables generalization across subjects and anatomical structures. \textbf{b,} Pelvic CT scans from lower body cadavers in the DeepFluoro dataset. \textbf{c,} Volume renderings of single-femur CTs in the Femur dataset. \textbf{d,} Maximum intensity projections (MIPs) of 3D rotational DSAs (rDSAs) from the Ljubljana dataset. \textbf{e-g,} After \qty{12}{\hour} of self-supervised training, \textit{de novo} networks (\blue) produce accurate initial pose estimates. However, finetuned models (\pink) match \textit{de novo} performance with only \qty{5}{\minute} of training. We report mean Target Registration Error (mTRE), which measures localization accuracy for 3D anatomical targets and is the most stringent pose estimation metric~(\cref{tab:metrics}). Other metrics are surveyed in~\cref{sec:metrics}. Error bars represent 90\% confidence intervals of pose estimation error across all patient X-rays. \textbf{h-j,} Synthetic X-rays rendered from poses predicted by various models after \qty{5}{\minute} of training (\textit{top}). Only finetuned models (\pink) achieve acceptable error at this stage, while \textit{de novo} models (\blue) achieve comparable accuracy after \qty{12}{\hour} of training. Ground truth and estimated fiducials are not used during pose estimation, but rather \textit{post hoc} to visualize and quantify registration error. In the DeepFluoro and Ljubljana datasets, fiducial markers were manually annotated in 3D, while fiducials in the Femur dataset were extracted from the femur surface on CT. In all datasets, 3D fiducials were projected into 2D using ground truth C-arm poses.}
    \label{fig:pretraining}
\end{figure*}

\section{Patient-specific networks in 5 minutes}
To assess our foundation model's performance and generalization to real X-ray images across multiple contexts, we evaluated \name on all publicly available human 2D/3D registration datasets with ground truth C-arm poses: (i) DeepFluoro~\cite{grupp2020automatic}, a collection of pelvic X-rays and CTs from six lower body cadavers, with six volumes and 362 X-rays (24--111 per subject); (ii) Femur~\cite{flepp2025automatic}, another cadaveric dataset with 49 distal femur X-rays from five subjects, each with an accompanying CT scan; (iii) Ljubljana~\cite{pernus20133d} containing 2D and 3D digital subtraction angiography (DSA) images from 10 endovascular neurosurgery patients. Each patient has one 3D rotational DSA (rDSA) and two 2D DSAs, as intraoperative images in endovascular procedures are commonly acquired using a biplane C-arm. Representative X-rays and volumes from each dataset are visualized in \cref{fig:wbct}a-c and \cref{fig:pretraining}b-d, respectively, alongside cropped scans from our whole-body pretraining corpus with similar anatomy (\cref{fig:pretraining}a). \cref{fig:wbct}d and \cref{fig:bch}a visualize internal clinical datasets of adult and neonatal CT/MR angiograms, respectively, which are used to further quantify the accuracy and scalability of \name.

To establish baseline performance, we first trained a \textit{de novo} (\ie from \uline{random initialization}) pose regression network for each subject using synthetic X-rays simulated from the subject's own volume. Each model used a ResNet34 backbone and was trained with a single NVIDIA RTX 6000 Ada GPU. \cref{fig:pretraining}e-g report the networks' average test error throughout the \qty{12}{\hour} training schedule (\blue curves) as mean Target Registration Error (mTRE) on real intraoperative X-rays. Simply applying \textit{de novo} models without further optimization successfully generalized to the subject's real X-rays, achieving a median mTRE of \qtyci{41.1}{\mm}{IQR}{29.6}{56.5} on DeepFluoro, \qtyci{55.5}{\mm}{IQR}{30.2}{90.2} on Femur, and \qtyci{44.4}{\mm}{IQR}{33.8}{60.4} on Ljubljana), thereby validating our renderer and self-supervised training strategy.

Next, we evaluated our foundation model's ability to rapidly adapt to new subjects using \uline{patient-specific finetuning}. Trained on subjects from the pretraining corpus with diverse clinical presentations~(\cref{fig:pretraining}a), our foundation model generalized to novel subjects in only \qty{5}{\minute} of finetuning, achieving comparable accuracy to the \textit{de novo} models, which were trained for \qty{12}{hrs} on the subject's own scan. Specifically, with \qty{5}{\minute} of finetuning, our models achieved median mTREs of \qtyci{53.6}{\mm}{IQR}{37.3}{75.2} on DeepFluoro, \qtyci{74.7}{\mm}{IQR}{50.8}{113.1} on Femur, and \qtyci{62.1}{\mm}{IQR}{35.8}{76.1} on Ljubljana. \cref{fig:pretraining}h-j illustrate initial C-arm poses estimated by our \textit{de novo} and finetuned models on sample X-rays. Given \qty{5}{\minute} of training time, the \textit{de novo} models are grossly inaccurate and converge slowly over hours, whereas the finetuned model's estimates are already accurate. Thus, training on multiple scans comprising diverse morphologies and pathologies enables our foundation model to build population-level representations that can be rapidly adapted to new subjects. However, \textit{de novo} subject-specific training with \name may be important for specialized scenarios, such as patients with highly distinct anatomy (\eg \textit{situs inversus}).

Unlike the controlled DeepFluoro and Femur cadaver studies, the Ljubljana dataset, acquired during live neurovascular interventions, presents two significant domain shifts. First, interventionalists routinely modify image acquisition parameters during procedures to enhance visualization of anatomical structures (\eg independently panning the C-arm detector or narrowing the field of view), resulting in unique intrinsic parameters for each X-ray. This violates a core assumption of our network architecture that synthetic X-rays are acquired with a canonical set of intrinsic parameters. To maintain compatibility with existing clinical workflows, we developed a geometric transform that resamples acquired X-rays to match the intrinsic parameters used during training~(\cref{fig:resample}), allowing use of the same network even as the interventionalist changes acquisition parameters. Second, publicly available rDSA data is limited compared to orthopedic imaging. To overcome this, our foundation model pretraining corpus includes high-resolution time-of-flight magnetic resonance angiograms (MRAs) from 61 healthy subjects~\cite{nitrc2017mra}, which we preprocessed to extract the neurovascular tree~(\cref{fig:pretraining}a, \textit{right})~\cite{xu2024vesselboost}. Despite these large domain shifts, our foundation model performed well on real 2D DSA images~(\cref{fig:pretraining}g), demonstrating that the pretraining dataset modality need not exactly match the clinically acquired scans, and that \name generalizes across diverse imaging conditions and anatomical targets in real-world clinical settings.

\begin{figure*}[htbp!]
    \centering
    \includegraphics[width=\linewidth]{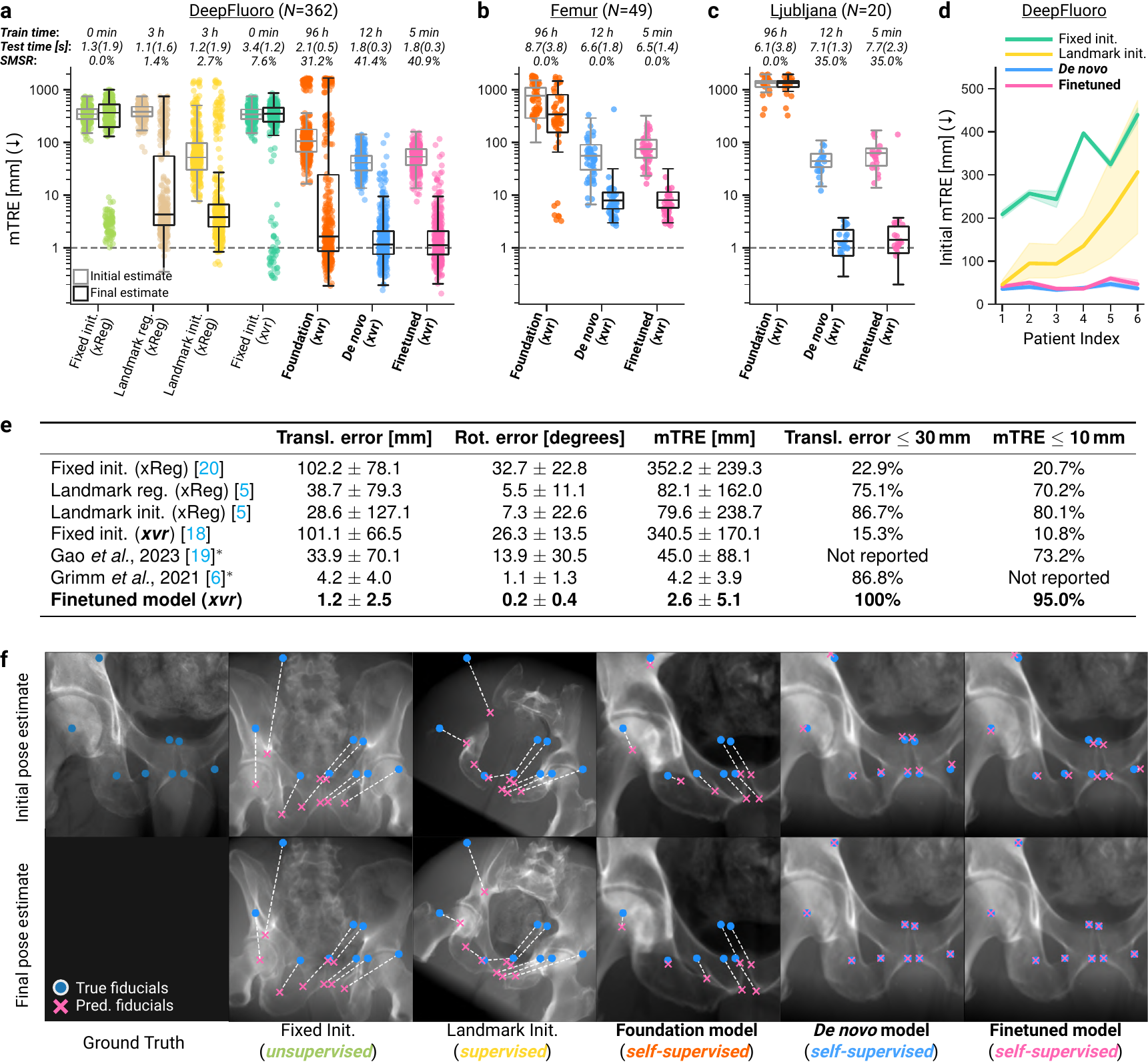}
    \caption{\textbf{Differentiable pose refinement achieves consistently accurate registration.} \textbf{a-c,} Initial and final pose estimate errors for multiple initialization and iterative pose refinement strategies. Each method is annotated with the neural network training times, iterative pose refinement times (mean and standard deviation), the percentage of X-rays that are successfully registered with less than \qty{1}{\mm} of error, and the renderer backend. \textbf{d,} Our patient-specific neural networks achieve low initial pose estimation errors across all patients in the DeepFluoro dataset, whereas supervised methods exhibit high inter-subject variation and frequent out-of-distribution failures as non-standard views are rare in the training set. Error bars represent 90\% confidence intervals of pose estimation error across all patient X-rays. \textbf{e,} Benchmark evaluations of 2D/3D registration methods on pelvic X-ray datasets. Asterisks ($^*$) indicate results reported in the original publications on similar but not identical data, necessitating nuanced interpretation: \citet{gao2023fully} also evaluate on the DeepFluoro dataset, but calculate mTRE using points on the surface of 3D bone segmentations, while \citet{grimm2021pose} evaluate a different, non-public pelvic dataset. \textbf{f,} Initial pose estimates produced by various pose estimation strategies for a particularly challenging intraoperative X-ray (\textit{top}). The extreme cranial angle of this view deviates substantially from a standard frontal view (\textit{Fixed Initialization}). Such poses are severely underrepresented in the training set of real X-ray images, and the supervised model (\textit{Landmark Initialization}) consequently suffers an out-of-distribution failure, predicting an implausible initial pose. In contrast, the \textit{de novo} and finetuned patient-specific models produce highly accurate initial pose estimates that are quickly refined to yield consistent millimeter-accurate registrations. Ground truth and estimated fiducial markers are used only for \textit{post hoc} error visualization and quantification, not for pose estimation.}
    \label{fig:benchmark}
\end{figure*}

\section{Robust intraoperative 2D/3D registration}

High-stakes interventions require intraoperative registration that is accurate to within a few millimeters. Therefore, we used iterative optimization to further refine the initial pose estimates produced by our finetuned networks, which differ from the ground truth C-arm poses in DeepFluoro, Femur, and Ljubljana by \qtyrange{30}{90}{\mm}. Using the differentiability of our renderer, we optimize the camera pose by maximizing the similarity between a real X-ray and a synthetic X-ray via automatic differentiation. \cref{fig:benchmark} reports the accuracy of \name versus previous 2D/3D registration methods, evaluating multiple strategies for initialization and refinement.

\subsection*{Initialization.}
Using the DeepFluoro dataset, we compare against existing pose estimation methods. We first compared our self-supervised pose regression networks to \textit{fixed initialization}~\cite{grupp2019pose}, where the same manually selected initial pose is used for every X-ray from all patients (\eg a standard frontal or lateral pose). This resulted in consistently high error (mTRE of \qtyci{342.7}{\mm}{IQR}{276.6}{426.2}), as clinicians frequently acquire non-standard views during interventions. Next, we evaluated \textit{landmark initialization}~\cite{grupp2020automatic}, a supervised deep learning method that trains a UNet~\cite{ronneberger2015u} to localize manually annotated landmarks in 2D X-rays, which are then used to estimate the C-arm pose with the Perspective-n-Point (PnP) algorithm~\cite{li2012robust}. We evaluated this method using leave-one-out cross-validation across the six subjects. \textit{Landmark initialization} achieved mTRE of \qtyci{52.1}{\mm}{IQR}{30.8}{98.5}, but also exhibited high inter-subject variability compared to \name. On the three most challenging subjects in DeepFluoro, our finetuned models achieved mean mTREs of \qty{49.3}{\mm}, \qty{39.3}{\mm}, and \qty{41.7}{\mm}, while \textit{landmark initialization} achieved \qty{122.2}{\mm}, \qty{233.4}{\mm}, and \qty{307.0}{\mm}, respectively~(\cref{fig:benchmark}d). \textit{Landmark initialization} and \textit{fixed initialization} fail on the same cases, suggesting that supervised learning also struggles with non-standard acquisitions. To demonstrate this, we visualized the initial pose estimates produced by various models for an unconventional X-ray~(\cref{fig:benchmark}f). As oblique acquisitions are underrepresented in the limited number of real X-rays available for supervised training, \textit{landmark initialization} suffers from an out-of-distribution failure. Our patient-specific framework is instead robust by design as it trains on ample non-standard synthetic X-rays exclusively generated from the patient undergoing the intervention.

Another disadvantage of (semi-)supervised landmark-based models is that their methodologies typically do not extend to novel anatomical structures. For example, in the Femur and Ljubljana datasets, there are no 2D segmentation masks from which to regress annotated fiducials, so landmark localization is not feasible with the UNet model proposed in~\cite{grupp2020automatic}. The neurovasculature in particular is highly heterogeneous, so much so that population-level landmark detection is infeasible~\cite{potente2017vascular}. Lastly, the small sample size of these datasets ($N=49$ and $N=20$ X-rays, respectively) makes it unlikely that supervised models trained on these data would generalize to novel patients. In contrast, our image-based pose regression approach extends directly to the Femur and Ljubljana datasets, with the finetuned models achieving final pose estimation errors of \qtyci{8.0}{\mm}{IQR}{5.7}{11.4} and \qtyci{1.4}{\mm}{IQR}{0.7}{2.2}, respectively~(\cref{fig:benchmark}b,c).

\subsection*{Refinement.}
For each method in \cref{fig:benchmark}a-c, we report the training time and the submillimeter success rate (SMSR), defined as the percentage of X-rays with mTRE less than \qty{1}{\mm} following pose refinement. On DeepFluoro, the most accurate methods were \name's iterative optimization initialized with either our finetuned or \textit{de novo} networks, achieving SMSRs of 40.9\% and 41.4\%, respectively. Instead, initializing iterative optimization with our foundation model \uline{without} using any patient-specific adaptation achieved 31.2\% SMSR after \qty{96}{\hour} of pretraining~(\cref{fig:benchmark}a), demonstrating that patient-agnostic pretraining can still be useful resource-constrained scenarios that do not allow for any patient-specific finetuning. In contrast, iterative optimization with differentiable rendering from a \textit{fixed initialization}~\cite{grupp2019pose} also does not require patient-specific training, but achieved an SMSR of only 7.6\%. 

To evaluate the utility of our differentiable renderer, we also compared against iterative optimization performed using xReg, a gradient-free optimization method~\cite{grupp2020automatic}. xReg did not achieve sufficient accuracy, producing SMSRs of just 0.3\% and 3.3\% for the \textit{fixed}~\cite{grupp2019pose} and \textit{landmark initializations}~\cite{grupp2020automatic}, respectively~(\cref{fig:benchmark}a). \textit{Landmark regularization}~\cite{grupp2020automatic}, an extension of xReg that uses predicted 2D landmarks as an additional loss term during optimization from the fixed initialization, achieved an SMSR of 1.7\%.

To further contextualize our performance, we compare against the reported empirical evaluations of two previous 2D/3D registration methods: (i) \citet{gao2023fully} use a neural network to learn a convex image similarity function for iterative optimization, and (ii) \citet{grimm2021pose} propose a semi-supervised method that transfers manually annotated keypoints from a training corpus to novel patients. Both papers evaluate their methods on real pelvic X-rays, with \citet{gao2023fully} also using the DeepFluoro dataset, however, neither framework provides code, models, and data splits. Therefore, we emphasize that these are indirect comparisons requiring nuanced interpretation. We evaluate \name using the same metrics used in the original publications (\cref{fig:benchmark}e), demonstrating that \name achieves the most accurate and reliable results of all tested methods on pelvic anatomy.

Our patient-specific finetuned (trained for \qty{5}{mins}) and \textit{de novo} models (trained for \qty{12}{hrs}) were not significantly different in accuracy from each other on any dataset (Bonferroni-corrected $p_{\mathrm{adj}} \geq 0.05$), confirming that rapid patient-specific adaptation of our foundation model achieves comparable accuracy to training from scratch on any specific patient, while requiring 144$\times$ less training time~(\cref{tab:ranks}; \cref{sec:stats}).

\begin{figure*}[hbt!]
    \centering
    \includegraphics[width=\linewidth]{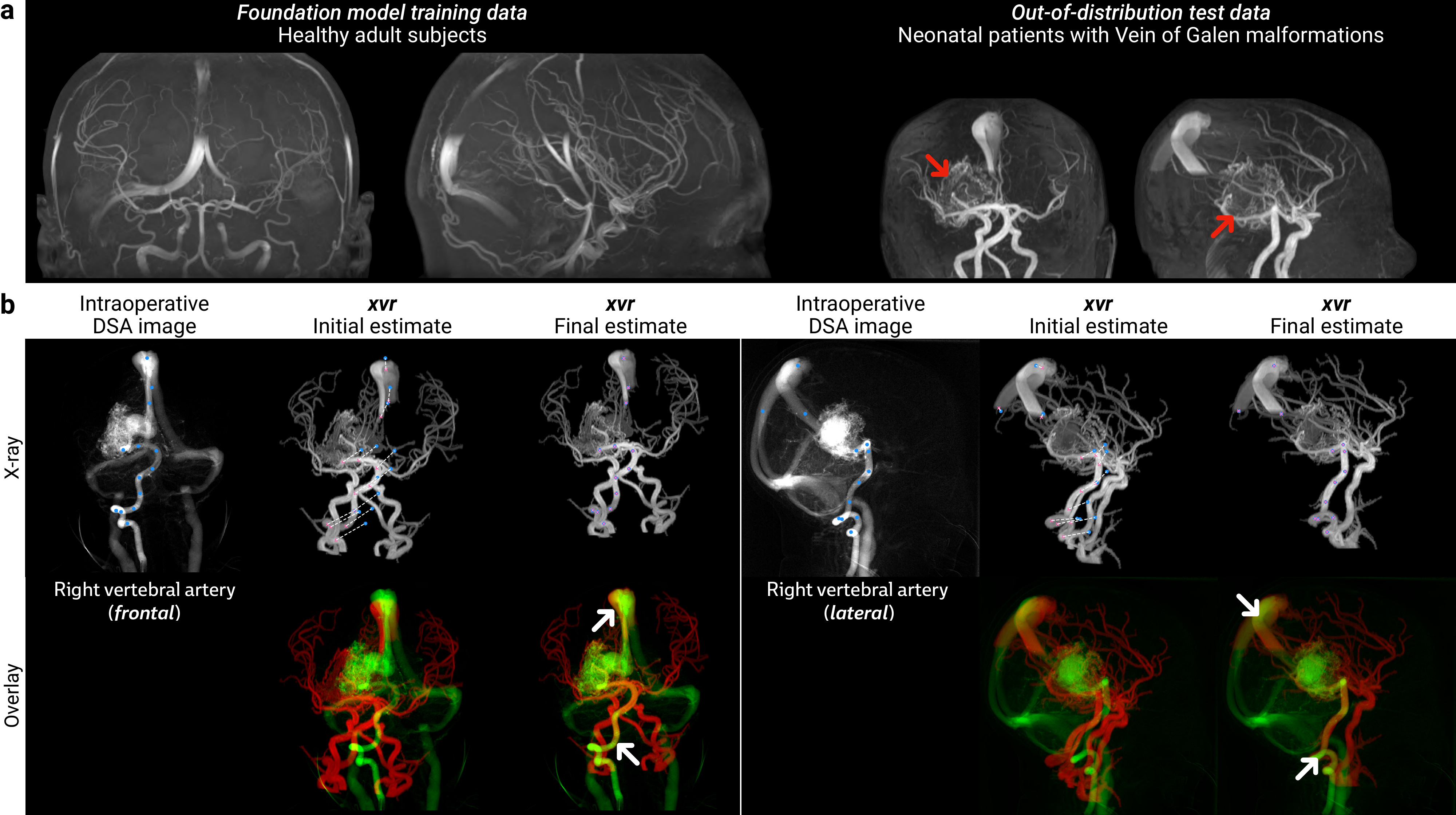}
    \caption{\textbf{Extreme out-of-distribution generalization.} \textbf{a,} Maximum intensity projections of 3D MRAs from a healthy adult subject in our pretraining corpus (\textit{left}) versus a neonatal patient with a large Vein of Galen malformation (\textit{right}, indicated by red arrows). \textbf{b,} Our foundation model finetuned on neonatal MRAs for \qty{5}{\minute} per patient ($N=4$) successfully registered intraoperative DSAs, achieving median mTRE \qtyci{2.3}{\mm}{IQR}{1.5}{3.0}, despite being pretrained exclusively on adult anatomy. Note that MRA captures only large arterial flow, while selective DSA injection reveals microvasculature and venous drainage absent from the preoperative scan, challenging registration, as corresponding structures may be invisible between modalities. White arrows indicate overlapping anatomical landmarks post registration.}
    \label{fig:bch}
    \vspace{-1em}
\end{figure*}

\begin{figure*}[htbp!]
    \centering
    \includegraphics[width=\linewidth]{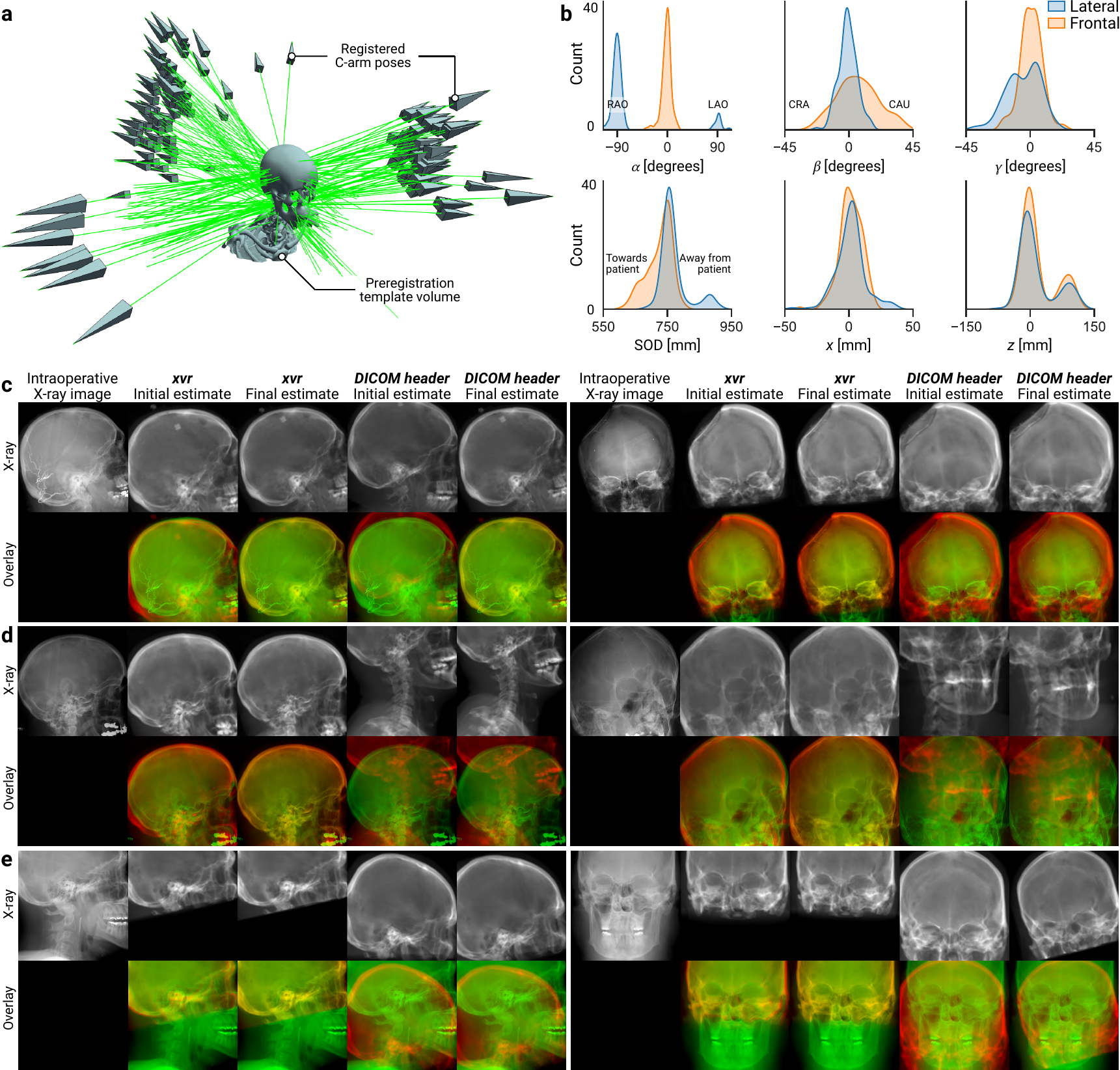}
    \caption{\textbf{\name enables the rapid registration of large amounts of real-world clinical data.} \textbf{a,} A patient-agnostic pose estimation model was trained using synthetic X-rays rendered from 61 preregistered head CTs in the TotalSegmentator dataset. Using this model and iterative pose refinement, 122 intraoperative X-rays from 50 neurosurgical patients at Brigham and Women's Hospital were registered to their corresponding preoperative 3D imaging. Registered C-arm poses from all 122 X-rays are visualized relative to the template head CT used for 3D registration. \textbf{b,} Distributions of estimated pose parameters reveal clinical patterns of interest, \eg right anterior oblique (RAO) lateral X-rays are acquired 8$\times$ more frequently in this dataset than left anterior oblique (LAO) X-rays. \textbf{c-e,} Our evaluations using manual annotations by trained neuroradiologists and neurointerventionalists revealed that \name achieved lower error (\qtyci{1.2}{\mm}{IQR}{0.5}{1.9}) than registrations initialized from pose parameters in the DICOM header (\qtyci{180.3}{\mm}{IQR}{106.2}{245.0}). \textbf{c,} \name's neural network retains its accuracy even on intraoperative images containing interventional findings, such as embolized vessels or craniotomies, not represented in the pretraining dataset. \textbf{d,} Pose parameters in the DICOM header do not account for the patient's positioning relative to the C-arm, often leading to insurmountably high initial pose estimation error. In contrast, \name produces consistently accurate initial pose estimates, even for non-standard views. \textbf{e,} Compared to CTs in benchmark datasets, clinical CTs sometimes image smaller fields-of-view to limit radiation exposure. Even with this limitation, \name can still register partial CTs to full field-of-view X-rays. Following registration, 3D information within the skull's rigid structure (\eg tumor boundaries or a road map to a lesion) can be reprojected from CT onto intraoperative X-rays.}
    \label{fig:brigham}
\end{figure*}

\section{Generalizing to real-world clinical data}

\subsection*{Extreme out-of-distribution generalization.}
To evaluate performance on populations and conditions completely unrepresented in our pretraining data, we patient-specific finetuned our foundation model on time-of-flight MRAs from four neonatal patients at Boston Children's Hospital presenting with Vein of Galen malformations. This cohort represents an extreme distribution shift from our pretraining data: pediatric patients are entirely absent from our adult-only corpus, and this vascular pathology is exceptionally rare (less than 1 in 50,000 live births~\cite{nunez2021epidemiology}). These patients also differ dramatically from healthy adults in size and exhibit pathological lesions~(\cref{fig:bch}a). Furthermore, the 2D DSAs also visualizes microvasculature and venous drainage entirely invisible on the 3D MRAs.  Despite this multi-faceted domain shift, \qty{5}{\minute} of patient-specific finetuning on average intensity projections simulated from preoperative 3D MRAs enabled accurate registration to intraoperative 2D DSA. The models achieved mTREs of \qtyci{2.3}{\mm}{IQR}{1.5}{3.0}, demonstrating that our approach generalizes reliably to patient populations and clinical contexts entirely absent from pretraining data~(\cref{fig:bch}b).

\subsection*{Scalability on large datasets.}
To further investigate \name in real-world clinical settings, we registered 50 CT angiograms (CTAs) and 122 DSAs from neurosurgery patients at Brigham and Women's Hospital \textit{\uline{without patient-specific adaptation}}. Using 61 prealigned head CTs from the TotalSegmentator dataset~\cite{wasserthal2023totalsegmentator}, we trained a patient-agnostic pose regression network for skull radiographs. For each DSA sequence, we estimated the C-arm pose from the first presubtraction frame, which highlights bony craniofacial structures useful for rigid registration. After rigidly registering each patient's CTA to the training template, we automatically corrected the network-estimated poses using our iterative optimizer.

To evaluate registration accuracy, we manually annotated corresponding keypoints on each X-ray and CT image pair and computed ground truth C-arm poses using the PnP algorithm~\cite{li2012robust}. \name achieved a median mTRE of \qtyci{1.2}{\mm}{IQR}{0.5}{1.9}, further demonstrating generalization to novel anatomical structures in real-world clinical data. Manual annotation of these keypoints required over 40 hours of clinicians' time, whereas pose estimation and refinement with \name ran completely automatically in just 10 minutes on a consumer-grade GPU (NVIDIA GeForce RTX 3060) at the clinical site. We visualize registered C-arm poses relative to the registration template volume~(\cref{fig:brigham}a) alongside distributions of the recovered pose parameters~(\cref{fig:brigham}b). For comparison, we also evaluated iterative optimization initialized from the C-arm pose encoded in the DICOM header~(\cref{sec:dicom}). Because these parameters do not account for the positioning of the patient relative to the C-arm, this approach often yielded inaccurate registrations (\qtyci{180.3}{\mm}{IQR}{106.2}{245.0}).

This clinical dataset also contained novel domain shifts between the 2D and 3D imaging. For example, some intraoperative DSAs captured surgical interventions such as embolized blood vessels or craniotomies~(\cref{fig:brigham}c). These findings were absent from preoperative volumes and therefore were not represented in the synthetic X-rays used for training or pose refinement. Despite this, our patient-agnostic model produced accurate initial and final pose estimates, highlighting \name's robustness to interventional changes. Additionally, many clinical CTAs imaged only a portion of the skull to minimize radiation exposure, producing partial synthetic X-rays that contain only a fraction of the anatomy visible in the real X-rays~(\cref{fig:brigham}e). Even with partial volumes, \name's pose refinement protocol produced accurate alignments.

\section{Discussion}
The registration of intraoperative 2D X-rays to 3D preoperative scans is a prerequisite for numerous advanced image guidance techniques, including vertebral level localization~\cite{abumoussa2023machine}, surgical plan reprojection~\cite{naik2022hybrid, metz2009patient}, instrument tracking~\cite{wagner20164d}, motion correction~\cite{huynh2020artificial}, and surgical robotics~\cite{kim2022telerobotic}. However, existing 2D/3D registration methods fail to achieve consistent and automatic performance across diverse anatomical structures, patient populations, and clinical specialties~\cite{unberath2021impact}. As such, interventions cannot integrate these tools due to the safety and robustness concerns~\cite{yip2023artificial, varghese2024artificial}. To address these challenges, we developed \name, a self-supervised machine learning framework that enables the rapid training of neural networks for X-ray to volume registration individualized to a patient's own anatomy. Our approach provides reliable, accurate, and anatomically generic 2D/3D registration across multiple medical specialties.

\paragraph{Solving the data bottleneck.}
Machine learning for medical imaging is limited by the scarcity of expert-labeled training data. These limitations are more severe in interventional applications, as intraoperative X-rays are rarely saved in the electronic medical record~\cite{varghese2024artificial, yip2023artificial}, and almost never with calibrated C-arm poses required for supervised training. Moreover, surgical cohorts typically present with non-standard anatomy (\eg fractures, implants, tumors, musculoskeletal degeneration, \etc), creating a scenario where data-limited supervised models often fail to generalize. \name addresses this bottleneck by generating synthetic training images from a patient's \textit{own} preoperative imaging via differentiable X-ray rendering. Networks trained exclusively on synthetic X-rays successfully generalize to real intraoperative images~(\cref{fig:pretraining}e-g), eliminating the need for manual landmark annotations or ground truth poses required by existing (semi-)supervised methods. By removing the need for annotation, \name presents a training framework that does not disrupt existing workflows.


\paragraph{Improving generalization through intentional overfitting}.
Supervised models can overfit to their training set, and this ``one-model-fits-all'' approach exhibits high inter-subject variability and often fails on out-of-distribution samples~(\cref{fig:benchmark}). Like supervised pose estimation models, our patient-specific models are also extremely overfit. However, \textit{this is intentional}: instead of overfitting an arbitrary training set, we design our models to overfit to \uline{the specific patient} undergoing the intervention. By learning the appearance and geometry of a patient's anatomy from synthetic X-rays, \name achieves consistently high 2D/3D registration accuracy, significantly outperforming existing methods~(\cref{fig:benchmark}). By employing a comprehensive data augmentation pipeline~(\cref{fig:supp_training}), \name is also robust to intraoperative domain shifts, such as patient repositioning or surgical devices~(\cref{fig:brigham}).

\paragraph{Training patient-specific neural networks in just \qty{5}{\minute}.} 
\name also addresses a long-standing limitation of patient-specific models: their extensive training time. By first pretraining a foundation model on whole-body CT datasets, \name reduces patient-specific training time from hours~\cite{gopalakrishnan2024intraoperative} to just \qty{5}{\minute}~(\cref{fig:pretraining}). Our experiments show that the proposed whole-body foundation model generalizes to multiple real X-ray datasets of the pelvis, femur, neurovasculature, and skull. This demonstrates that multi-patient simulation enables pose regression networks to learn a population-level understanding of human anatomy~(\cref{fig:benchmark}), supporting rapid finetuning on new subjects. This approach is robust to significant domain shifts. For example, abdominal scans in our pretraining corpus comprise the entire pelvis, while the DeepFluoro cadaver volumes do not contain the top half of the torso~(\cref{fig:pretraining}b). Additionally, the time-of-flight MRAs in our corpus, in addition to being a completely different modality, are of healthy volunteers, whereas 3D rDSAs in Ljubljana contain vascular malformations~(\cref{fig:pretraining}d). 

\paragraph{Limitations and future work.} 

While our study is the largest and most anatomically diverse evaluation of 2D/3D registration algorithms on real X-ray images and MR/CT volumes to date, our overall sample size remains relatively small. We evaluate \name on 21 subjects from public datasets and 54 from private clinical sources, which reflects the paucity of openly available paired X-ray/CT data and the high cost of acquiring such images for evaluation. Although \name is statistically significantly better than prior work across heterogeneous datasets, these claims of out-performance should still be interpreted with caution. Stronger evidence of robustness will require larger cohorts spanning additional anatomical regions, disease presentations, and imaging protocols, alongside pilot studies prospectively integrating \name into intraoperative settings.

Furthermore, \name achieves consistent registration accuracy through iterative pose refinement, which is necessary because initial model pose estimates differ from the ground truth by 30--\qty{90}{\mm}, but currently adds 1--8 seconds of lag. Although iterative solvers currently outperform deep learning estimators in multiple image registration tasks~\cite{jena2024adeep, dey2025learning}, improving the accuracy of deep learning estimators is desirable as they yield predictions in milliseconds. Accurate \textit{real-time} pose estimation without iterative refinement would be required to power self-driving image-guidance systems~\cite{gagoski2022automated}. Beyond improving initial pose accuracy, our framework is readily extensible to several important 2D/3D registration scenarios. First, the pose regression network could be adapted to predict poses of multiple rigid bodies simultaneously, enabling the estimation of polyrigid transformations~\cite{arsigny2009fast, gopalakrishnan2026polypose}. Second, \name extends directly to multiview rigid registration by incorporating a differentiable cone-beam CT reconstruction operator and volumetric loss term during optimization. Third, joint estimation of the intrinsic and extrinsic camera parameters would enhance robustness in atypical scenarios where DICOM headers may be inaccessible or unreliable~(\cref{fig:supp_intrinsics}).

In summary, \name achieves consistently accurate patient-specific 2D/3D registration through self-supervised learning on synthetic X-rays. Evaluated across several clinical datasets, our framework demonstrates robust performance across diverse anatomies, pathologies, and acquisition protocols. By combining differentiable rendering with rapid pretrained model finetuning, \name provides a foundation for next-generation image guidance. The framework is freely available at \url{https://xvr.csail.mit.edu}.

\putbib
\end{bibunit}

\clearpage
\begin{bibunit}
\methods
\section{Differentiable X-ray rendering}
\label[meth]{sec:diffdrr}

\paragraph{Geometry of the 3D volume.}
The physical spacing and orientation of the 3D volume is determined by its affine matrix $\mathbf A$, which maps voxel coordinates $(i, j, k) \in \mathbb N^3$ to world coordinates $(x, y, z) \in \mathbb R^3$, represented in homogeneous coordinates:
\begin{equation}
    \mathbf A = \begin{bmatrix}
        \Delta x & 0 & 0 & O_x \\    
        0 & \Delta y & 0 & O_y \\    
        0 & 0 & \Delta z & O_z \\    
        0 & 0 & 0 & 1
    \end{bmatrix} \,,
\end{equation}
where $\mathbf O = (O_x, O_y, O_z)$ is the origin of the 3D volume in world coordinates and $\mathbf \Delta = (\Delta x, \Delta y, \Delta z)$ is the voxel dimensions with units of millimeters per voxel. The signs of the elements in $\Delta$ determine the orientation of the 3D volume along each of the axes. If the number of pixels in each dimension of the 3D is $\mathbf N = (N_x, N_y, N_z)$, then the patient's isocenter in world coordinates is
\begin{equation}
    \mathbf i = \frac{\mathbf N \odot \mathbf \Delta}{2} + \mathbf O \,,
\end{equation}
where $\odot$ is the Hadamard product, representing element-wise multiplication.

\subsection*{Geometry of the C-arm.}
We follow the standard approach of modeling a C-arm as a pinhole camera and express the X-ray image formation model using projective geometry. The intrinsic matrix $\mathbf K$, which maps camera coordinates to pixel coordinates~\cite{hartley2003multiple}, can be factored as
\begin{align}
    \mathbf K &= 
    \begin{bmatrix}
        \nicefrac{1}{s_x} & 0 & \nicefrac{W}{2} \\
        0 & \nicefrac{1}{s_y} & \nicefrac{H}{2} \\
        0 & 0 & 1
    \end{bmatrix}
    \begin{bmatrix}
        f & 0 & o_x \\
        0 & f & o_y \\
        0 & 0 & 1 
    \end{bmatrix}
    = \pKi \iKc \,,
\end{align}
where $f$ is the C-arm's source-to-detector distance (\ie the focal length) in millimeters, $(o_x, o_y)$ is the optical center of the C-arm in millimeters, $(s_x, s_y)$ are the spacings of pixels in the detector plane with units of millimeters per pixel, and $(H, W)$ are the height and width in pixels of the detector plane. Matrix $\iKc$ maps camera coordinates to image coordinates (with units of millimeters) and $\pKi$ maps image coordinates to pixel coordinates. 

The X-ray source is initialized at the origin in world coordinates $(0, 0, 0)$ and the center of the detector plane is initialized at $(0, 0, f)$, where the intrinsic parameters in $\mathbf K$ determine the initial positions of the pixel centers in the detector plane. These initial positions are then reoriented such that the depth dimension of the renderer is aligned with either the posterior-anterior or anterior-posterior dimension of the CT scan (\ie the $y$-axis).
The positions of the X-ray source and detector can be reoriented using any 3D rigid transformation $\mathbf T \in \SE3$, the special Euclidean group. Specifically, $\mathbf T$ comprises a 3D rotation $\mathbf R \in \SO3$ and a 3D translation $\mathbf t \in \mathbb R^3$, which we expressed in homogeneous coordinates as
\begin{equation}
    \label{eq:camera-to-world}
    \wTc = 
    \begin{bmatrix}
        \mathbf R & \mathbf{Rt} \\
        \mathbf 0^T & 1
    \end{bmatrix} \,.
\end{equation}
That is, $\mathbf T$ determines the geometry of the C-arm by first translating the X-ray source and detector by $\mathbf t$, then rotating the camera's coordinate frame by $\mathbf R$. This transformation, referred to in this paper as the C-arm pose or more generally as the camera-to-world matrix, and the intrinsic matrix $\mathbf K$ together define the projection matrix
\begin{equation}
    \label{eq:perspective-projection}
    \mathbf \Pi = \mathbf K [\mathbf R^T \mid -\mathbf{t}] \,,
\end{equation}
which maps any point in world coordinates to pixel coordinates using a perspective projection. In \cref{eq:perspective-projection}, points in world coordinates are first mapped to camera coordinates with the world-to-camera matrix:
\begin{equation}
    \cTw = \mathrm{inv}(\wTc) = 
    \begin{bmatrix}
        \mathbf R^T & -\mathbf t \\
        \mathbf 0^T & 1 
    \end{bmatrix} \,.
\end{equation}

There are many choices of parameterization for the rotation matrix $\mathbf R$, such as Euler angles, quaternions, the axis-angle parameterization, the tangent space $\so3$, \etc. \name supports all of these parameterizations, whether for specifying the C-arm pose or performing gradient-based pose optimization. Commercial C-arms define rotation matrices using Euler angles with the convention
\begin{equation}
    \label{eq:c-arm-rot}
    \mathbf R(\alpha, \beta, \gamma) = \mathbf R_z(\alpha) \mathbf R_x(\beta) \mathbf R_y(\gamma) \,,
\end{equation}
where $\mathbf R_i$ is a $3 \times 3$ matrix denoting rotation about the $i$-axis for $i \in \{x, y, z\}$. Here, $\alpha$ refers to the left-right anterior oblique rotational axis (LAO/RAO) and $\beta$ refers to the cranial-caudal rotational axis (CRA/CAU). The geometric convention of C-arms as implemented in \name is illustrated in \cref{fig:overview}C.

\begin{figure*}[t!]
    \centering
    \includegraphics[width=\linewidth]{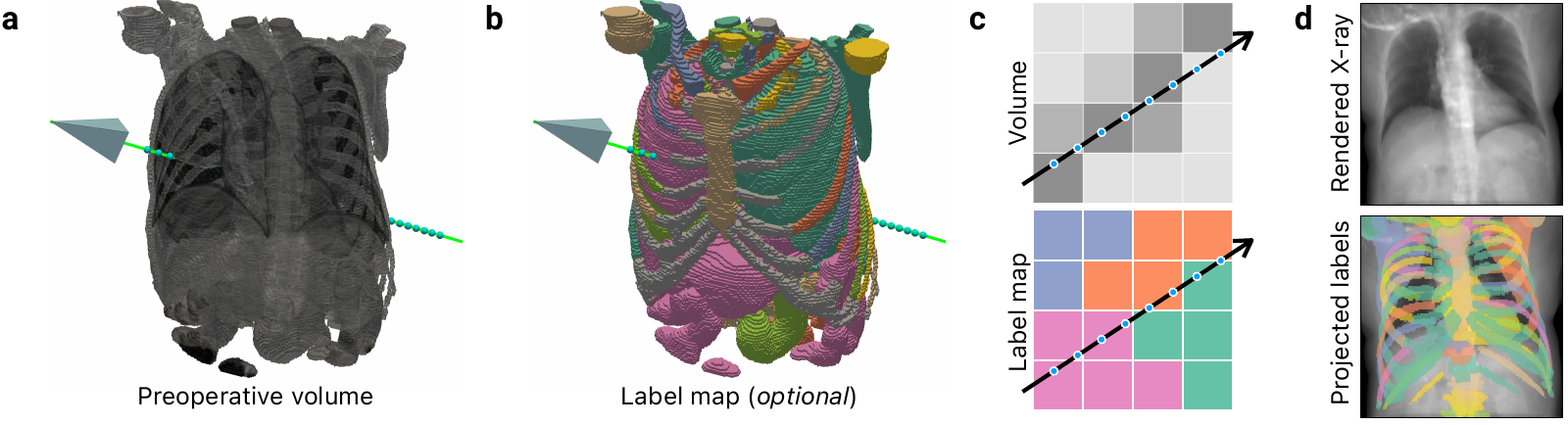}
    \caption{\textbf{\name implements a physics-based differentiable renderer that simulates the geometry of an X-ray C-arm to generate photorealistic X-ray images from 3D volumes.} \textbf{a,} Our renderer requires two inputs: a 3D volume from which to generate synthetic X-rays and the pose of the C-arm (represented with a camera frustum). Our renderer is differentiable with respect to the C-arm pose, enabling us to use gradient-based optimization to register X-ray images to 3D volumes. \textbf{b,} Optionally, a 3D label map of the preoperative volume can also be used to render X-rays of specific anatomical structures. \textbf{c,} A pictorial overview of trilinear interpolation, one of the ray tracing methods we implemented to render synthetic X-rays, along with Siddon's method~\cite{siddon1985fast}. \textbf{d,} In addition to developing fully differentiable implementations of ray tracing with trilinear interpolation and Siddon's method, we also adapt these algorithms to project 3D anatomical labels onto 2D space, enabling structure-specific registration.}
    \label{fig:supp_renderer}
    \vspace{-1em}
\end{figure*}

\paragraph{Rendering equation.}
We derive a first-order model of X-ray image formation in a continuous form to inspire discretized computational implementations. Let $\mathbf s \in \mathbb R^3$ be the radiation point source and $\mathbf p \in \mathbb R^3$ be the location of a pixel on the detector plane, both defined in world coordinates. These points define the ray $\vec{\mathbf r}(\alpha) = \mathbf s + \alpha (\mathbf p - \mathbf s)$ for $\alpha \in [0, 1]$. This ray is cast through a heterogeneous medium (\eg human body) $\mathbf V : \mathbb R^3 \mapsto [0, \infty)$, where $\mathbf V(\mathbf x)$ represents the linear attenuation coefficient (LAC) at each point in the medium $\mathbf x \in \mathbb R^3$. Then, the pixel intensity induced by this ray is given by the Beer-Lambert law~\cite{swinehart1962beer}:
\begin{align}
    I_{\mathrm{BL}}(\vec{\mathbf r}) 
    &\triangleq I_0 \exp \left( -\int_{\mathbf x \in \mathbf{{\vec r}}} \mathbf V(\mathbf x) \mathrm d \mathbf x \right) \\
    &= I_0 \exp \left( -\int_0^1 \mathbf V\big(\vec{\mathbf r}(\alpha)\big) \| \dot{\vec{\mathbf r}}(\alpha) \| \mathrm d \alpha \right) \\
    &= I_0 \exp \left( -\| \mathbf p - \mathbf s \| \int_0^1 \mathbf V\big( \mathbf s + \alpha (\mathbf p - \mathbf s) \big) \mathrm d \alpha \right) \,,
\end{align}
where $I_0$ is the initial intensity of the X-ray beam. Expressing the log-transformed version of this quantity
\begin{align}
    I(\vec{\mathbf r}) 
    &\triangleq \log I_0 - \log I_{\mathrm{BL}}(\vec{\mathbf r}) \\
    &= \| \mathbf p - \mathbf s \| \int_0^1 \mathbf V\big( \mathbf s + \alpha (\mathbf p - \mathbf s) \big) \mathrm d \alpha \,.
    \label{eq:continuous-forward-model}
\end{align}
simplifies the formulation.

When rendering synthetic X-rays, we do not have access to the continuous form of $\mathbf V$, but a discrete analog from a preoperative 3D CT or MR volume. Therefore, we require numerical methods to analytically compute the integral in \cref{eq:continuous-forward-model}. The first integration technique we consider is Siddon's method~\cite{siddon1985fast}, which exactly computes a discretized version of \cref{eq:continuous-forward-model} as the sum of the linear attenuation coefficient in each voxel on the path of $\vec{\mathbf r}$, weighted by the intersection length of $\vec{\mathbf r}$ with each voxel:
\begin{equation}
    \label{eq:siddon}
    I(\vec{\mathbf r}) = \| \mathbf p - \mathbf s \| \sum_{m=1}^{M-1} \mathbf V \left [ \mathbf s + \frac{\alpha_{m+1} + \alpha_m}{2} (\mathbf p - \mathbf s) \right] (\alpha_{m+1} - \alpha_m) \,,
\end{equation}
where $\{\alpha_1, \dots, \alpha_M\}$ parameterizes the intersections of $\vec{\mathbf r}$ with the parallel planes comprising $\mathbf V$ as determined by the matrix $\mathbf A$. Additionally, $\mathbf V[\cdot]$ is an indexing operation that returns the linear attenuation coefficient of the intersected voxel (\ie nearest-neighbor interpolation). We have previously shown that \cref{eq:siddon} can be implemented in a completely differentiable manner~\citep{gopalakrishnan2022fast}. 

Instead of computing every plane intersection, which scales cubically with the resolution of $\mathbf V$, we can approximate the Beer-Lambert law using interpolatory quadrature. Thus, the second integration technique we consider uses trilinear interpolation to construct an estimate
\begin{equation}
    \label{eq:trilinear}
    I(\vec{\mathbf r}) \approx \frac{\| \mathbf p - \mathbf s \|}{M-1} \sum_{m=1}^{M-1} \mathbf V \left[ \mathbf s + \alpha_m (\mathbf p - \mathbf s) \right] \,,
\end{equation}
where $\{\alpha_1, \dots, \alpha_M\}$ parameterize $M$ evenly spaced points along $\vec{\mathbf r}$ and $\mathbf V[\cdot]$ represents trilinear interpolation. Trilinear interpolation is linear in $M$ and thus faster and less computationally expensive than Siddon's method. Therefore, to increase batch sizes for neural network training and decrease rendering time for iterative pose refinement, we use trilinear interpolation for all tasks that require rendering synthetic X-rays in \name by default.

Given an X-ray source $\mathbf s \in \mathbb R^3$ and a set of target pixels on the detector grid $\mathbf P \in \mathbb R^{n \times 3}$, we can reorient $\mathbf s$ and $\mathbf P$ by a rigid transform $\mathbf T$ to render an X-ray from any particular view. Therefore, we denote the rendered image by $\mathbf I = \mathcal P(\mathbf T) [\mathbf V]$, where $\mathcal P$ is the projection operator in \cref{eq:perspective-projection}. We implement both~\cref{eq:siddon} and~\cref{eq:trilinear} in PyTorch, making these forward models differentiable with respect to the input pose $\mathbf T$.

\begin{table*}[t!]
\resizebox{\textwidth}{!}{%
\begin{tabular}{@{}llccccccc@{}}
\toprule
 & & \multicolumn{4}{c}{\textbf{\name}} & \multicolumn{3}{c}{\textbf{xReg}} \\
\cmidrule(lr){3-6}\cmidrule(lr){7-9}
 & & Finetuned & \textit{De novo} & Foundation & Fixed init.~\cite{gopalakrishnan2022fast} & Fixed init.~\cite{grupp2019pose} & Landmark reg.~\cite{grupp2020automatic} & Landmark init.~\cite{grupp2020automatic} \\
\midrule
\multirow{8}{*}[-0.75em]{\rotatebox{90}{\textbf{DeepFluoro}}}
 & \multirow{2}{*}{\makebox[3cm][l]{mPE (2D)\hfill[mm]}}    
   & \textbf{0.2(0.2)} & \textbf{0.2(0.2)} & 0.2(3.0)      & 74.5(60.4)   & 86.3(80.2)   & 1.0(3.2)   & 0.9(0.4)  \\
 & & \textbf{93.9\%}   & 93.4\%            & 71.0\%        & 10.8\%       & 18.2\%       & 53.6\%     & 59.9\%    \\ \cmidrule(l){2-9}
 & \multirow{2}{*}{\makebox[3cm][l]{mRPE (2.5D)\hfill[mm]}} 
   & \textbf{0.6(0.8)} & 0.6(0.9)          & 0.9(16.5)     & 137.6(83.9)  & 151.3(133.0) & 2.2(29.6)  & 1.9(2.2)  \\
 & & 69.1\%            & \textbf{70.2\%}   & 52.5\%        & 9.7\%        & 8.8\%        & 16.9\%     & 20.2\%    \\ \cmidrule(l){2-9}
 & \multirow{2}{*}{\makebox[3cm][l]{mTRE (3D)\hfill[mm]}}   
   & \textbf{0.9(1.1)} & \textbf{0.9(1.1)} & 1.3(19.0)     & 277.4(137.1) & 335.4(249.1) & 3.1(39.9)  & 2.8(2.7)  \\
 & & 40.9\%            & \textbf{41.4\%}   & 31.2\%        & 7.6\%        & 0.0\%        & 1.4\%      & 2.7\%     \\ \cmidrule(l){2-9}
 & \multirow{2}{*}{\makebox[3cm][l]{dGeo (3D)\hfill[mm]}}   
   & \textbf{1.1(1.3)} & 1.2(1.3)          & 1.6(23.4)     & 352.8(203.3) & 365.9(329.5) & 4.3(52.3)  & 3.8(4.1)  \\
 & & 53.3\%            & \textbf{55.5\%}   & 41.7\%        & 8.0\%        & 0.0\%        & 3.3\%      & 6.0\%     \\ \midrule

\multirow{8}{*}[-0.5em]{\rotatebox[origin=c]{90}{\textbf{Femur}}}
 & \multirow{2}{*}{\makebox[3cm][l]{mPE (2D)\hfill[mm]}}
   & \textbf{2.0(0.9)} & 2.0(0.9) & 82.8(178.2) & \multicolumn{4}{c}{\multirow{8}{*}[-0.5em]{N/A}} \\
 & & \textbf{6.1\%} & \textbf{6.1\%} & 2.0\% & \multicolumn{4}{c}{} \\ \cmidrule(l){2-5}
 & \multirow{2}{*}{\makebox[3cm][l]{mRPE (2.5D)\hfill[mm]}}
   & 4.2(2.3) & \textbf{4.1(2.3)} & 198.6(246.2) & \multicolumn{4}{c}{} \\
 & & 0.0\% & 0.0\% & 0.0\% & \multicolumn{4}{c}{} \\ \cmidrule(l){2-5}
 & \multirow{2}{*}{\makebox[3cm][l]{mTRE (3D)\hfill[mm]}}
   & \textbf{8.0(5.7)} & 8.0(5.8) & 343.1(647.4) & \multicolumn{4}{c}{} \\
 & & 0.0\% & 0.0\% & 0.0\% & \multicolumn{4}{c}{} \\ \cmidrule(l){2-5}
 & \multirow{2}{*}{\makebox[3cm][l]{dGeo (3D)\hfill[mm]}}
   & 5.9(4.1) & \textbf{5.9(4.0)} & 277.9(421.8) & \multicolumn{4}{c}{} \\
 & & 0.0\% & 0.0\% & 0.0\% & \multicolumn{4}{c}{} \\ \midrule

\multirow{8}{*}[-0.75em]{\rotatebox{90}{\textbf{Ljubljana}}}
 & \multirow{2}{*}{\makebox[3cm][l]{mPE (2D)\hfill[mm]}}    
   & \textbf{0.7(0.1)} & \textbf{0.7(0.1)} & 394.7(660.5) & \multicolumn{4}{c}{\multirow{8}{*}{N/A}} \\
 & & 95.0\% & \textbf{100.0\%} & 0.0\% & \multicolumn{4}{c}{} \\ \cmidrule(l){2-5}
 & \multirow{2}{*}{\makebox[3cm][l]{mRPE (2.5D)\hfill[mm]}} 
   & \textbf{1.0(0.9)} & 1.0(1.0) & 826.7(576.2) & \multicolumn{4}{c}{} \\
 & & 45.0\% & \textbf{55.0\%} & 0.0\% & \multicolumn{4}{c}{} \\ \cmidrule(l){2-5}
 & \multirow{2}{*}{\makebox[3cm][l]{mTRE (3D)\hfill[mm]}}   
   & \textbf{1.4(1.5)} & \textbf{1.4(1.5)} & 1315.5(392.1) & \multicolumn{4}{c}{} \\
 & & \textbf{35.0\%} & \textbf{35.0\%} & 0.0\% & \multicolumn{4}{c}{} \\ \cmidrule(l){2-5}
 & \multirow{2}{*}{\makebox[3cm][l]{dGeo (3D)\hfill[mm]}}   
   & 1.2(1.3) & \textbf{1.2(1.1)} & 1384.1(719.1) & \multicolumn{4}{c}{} \\
 & & \textbf{20.0\%} & \textbf{20.0\%} & 0.0\% & \multicolumn{4}{c}{} \\ \bottomrule
\end{tabular}%
}
\caption{Pose estimation error for 2D/3D registration error metrics reported as the median and interquartile range error (mm) and submillimeter success rate (\%). The best-performing 2D/3D registration method for each metric is bolded. Mean target registration error (mTRE), the most stringent metric, is reported throughout the main text. Methods with the xReg backend cannot be evaluated on the Femur and Ljubljana datasets because they require manual segmentations of 2D X-rays from which to regress annotated fiducials.}
\vspace{-1em}
\label{tab:metrics}
\end{table*}

\paragraph{Considerations.} Our renderer does not model various higher-order effects such as scattering, polychromaticity, or the energy-dependent tissue responses. While our model is sufficient for 2D/3D registration, other machine learning tasks (\eg classification) trained using synthetic X-ray images may benefit from a more physically accurate, albeit non-differentiable, renderer such as DeepDRR~\cite{unberath2018deepdrr}.

\section{Pose estimation error metrics}
\label[meth]{sec:metrics}
Many metrics exist to assess the accuracy of 2D/3D registration results. Here, we derive previously proposed 2D/3D registration metrics and use them to evaluate the DeepFluoro, Femur, and Ljubljana datasets~(\cref{tab:metrics}). Throughout the main text, we report the mean Target Registration Error (mTRE) as it is the most stringent of the metrics we consider. 

\paragraph{Preliminaries.}
Let $\mathbf T, \mathbf{\hat T} \in \mathbf{SE}(3)$ be a ground truth and estimated C-arm pose, respectively. Let $\mathbf K$ be the C-arm's known intrinsic matrix, which we leverage to construct the projection matrices $\mathbf \Pi, \mathbf{\hat \Pi}$ in \cref{eq:perspective-projection}. Finally, let $\mathbf X \in \mathbb R^{3 \times M}$ be a set of $M$ fiducial markers annotated for every volume. In projective geometry, 3D points are typically represented using homogeneous coordinates in order to represent perspective projections with a single matrix operation~\cite{hartley2003multiple}. Specifically, we use $\pi: \mathbb R^3 \to \mathbb R^2$ to represent the nonlinear projection operation that maps $\begin{bmatrix}x & y & z\end{bmatrix}^T = \mathbf\Pi\begin{bmatrix}X & Y & Z & 1\end{bmatrix}^T$ to $\begin{bmatrix} x / z & y / z \end{bmatrix}^T \in \mathbb R^2$.

\noindent
\textbf{Mean Projection Error (mPE)} measures the distance between fiducials when projected onto the ground truth and estimated detector planes (\ie in 2D), respectively:
\begin{equation}
    \label{eq:mpe}
    \mathcal L_{\mathrm{mPE}}(\mathbf T, \mathbf{\hat T)} = \frac{1}{M} \| \pi(\mathbf X) - \hat\pi(\mathbf X) \|_2 \,.
\end{equation}

\noindent
\textbf{Mean Reprojection Error (mRPE)} lifts 2D projected fiducials onto a 2D surface embedded in 3D, enabling measurement of the distance between the detector planes (\ie in 2.5D):
\begin{equation}
    \label{eq:mrpe}
    \mathcal L_{\mathrm{mRPE}}(\mathbf T, \mathbf{\hat T}) = \frac{1}{M} \| f\mathbf K^{-1}(\pi(\mathbf X) - \hat\pi(\mathbf X)) \|_2 \,,
\end{equation}
where $f$ is the focal length of the C-arm, derived from $\mathbf K$.

\noindent
\textbf{Mean Target Registration Error (mTRE)} directly measures the distance between fiducial markers in world coordinates, ignoring the projection (\ie in 3D):
\begin{equation}
    \label{eq:mtre}
    \mathcal L_{\mathrm{mTRE}}(\mathbf T, \mathbf{\hat T}) = \frac{1}{M} \| \mathbf T(\mathbf X) - \mathbf{\hat T}(\mathbf X) \|_2 \,.
\end{equation}

\noindent
\textbf{Double geodesic distance} factors the distance between two rigid transforms into rotational and translational distances:
\begin{align}
    \mathcal L_\mathrm{rot}(\mathbf R, \mathbf{\hat R}) &= \arccos\left( \frac{\mathrm{tr}(\mathbf R^T \mathbf{\hat R}) - 1}{2} \right) \label{eq:drot} \\
    \mathcal L_\mathrm{arc}(\mathbf R, \mathbf{\hat R}) &= \frac{f}{2} \mathcal L_\mathrm{rot}(\mathbf R, \mathbf{\hat R}) \\
    \mathcal L_\mathrm{xyz}(\mathbf t, \mathbf{\hat t}) &= \| \mathbf t - \mathbf{\hat t} \|_2 \label{eq:dxyz} \,,
\end{align}
where multiplying \cref{eq:drot} by the radius $f/2$ converts the arc length from units of radians to millimeters. Finally, these metrics are combined into a single distance metric:
\begin{equation}
    \label{eq:dgeo}
    \mathcal L_{\mathrm{dGeo}}(\mathbf T, \mathbf{\hat T}) = \sqrt{\mathcal L_\mathrm{arc}(\mathbf R, \mathbf{\hat R}; f)^2 + \mathcal L_\mathrm{xyz}(\mathbf t, \mathbf{\hat t})^2} \,.
\end{equation}
Note that, unlike the other three metrics described in this section, $\mathcal L_\mathrm{dGeo}$ does not require manually annotated fiducials. Therefore, we use \cref{eq:dgeo} as a loss function when training pose regression neural networks in \name (see~\cref{eq:loss}).

\begin{figure*}[bp]
    \centering
    \includegraphics[width=\linewidth]{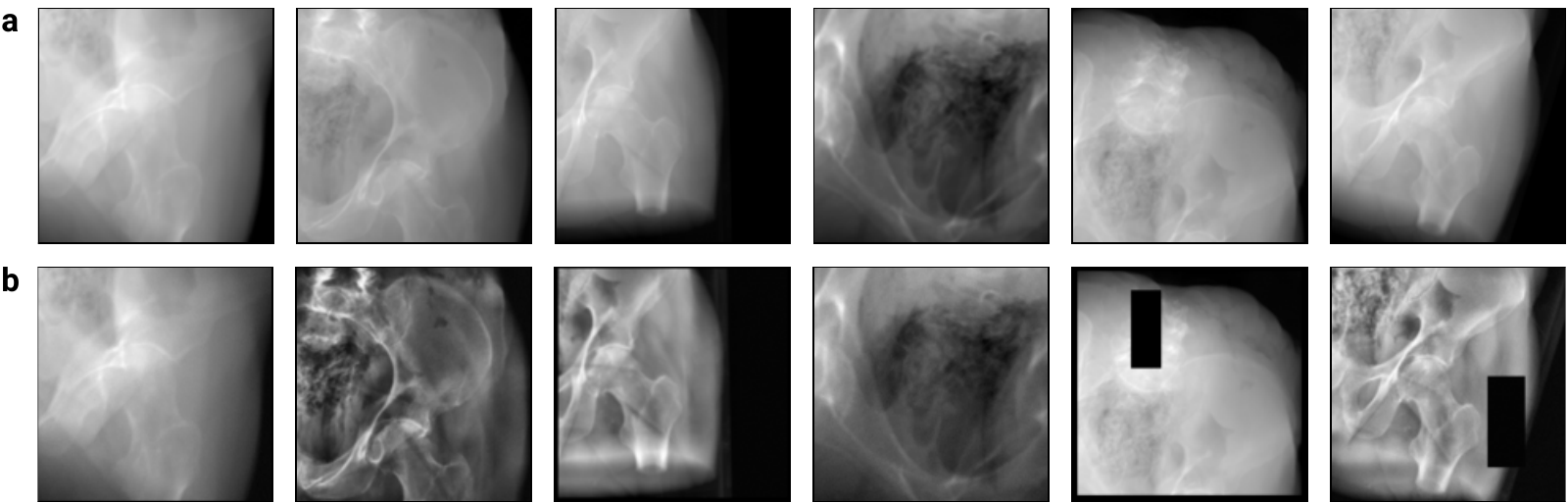}
    \caption{\textbf{Data augmentation pipeline to diversify synthetic X-ray training data.} \textbf{a,} Example synthetic X-rays rendered at random C-arm poses from the preoperative imaging. \textbf{b,} To improve the robustness of the pose estimation network, we perform data augmentations such as adding Gaussian noise, applying local contrast enhancement, masking rectangular patches, or simulated collimation.}
    \label{fig:supp_training}
\end{figure*}

\begin{table*}[t!]
\centering
\resizebox{0.7875\textwidth}{!}{
\begin{tabular}{@{}lcccccccccccc@{}}
\toprule
 & \multicolumn{6}{c}{Rotations [$^\circ$]} & \multicolumn{6}{c}{Translations [mm]} \\ \cmidrule(l){2-7} \cmidrule(l){8-13}
 & $\alpha_{\mathrm{min}}$ & $\alpha_{\mathrm{max}}$ & $\beta_{\mathrm{min}}$ & $\beta_{\mathrm{max}}$ & $\gamma_{\mathrm{min}}$ & $\gamma_{\mathrm{max}}$ & $x_{\mathrm{min}}$ & $x_{\mathrm{max}}$ & $y_{\mathrm{min}}$ & $y_{\mathrm{max}}$ & $z_{\mathrm{min}}$ & $z_{\mathrm{max}}$ \\ \midrule
Foundation & {$-$180}            & 180  & {$-$60} & 60  & {$-$30} & 30  & {$-$150} & 150  & 450  & 1000 & {$-$150} & 150  \\
DeepFluoro & {\phantom{$-$}135}  & 225  & {$-$45} & 45  & {$-$15} & 15  & {$-$150} & 150  & 450  & 1000 & {$-$150} & 150  \\
Femur      & {\phantom{$-$}\,75} & 270  & {$-$20} & 20  & {$-$20} & 20  & {$-$75}  & 75   & 650  & 950  & {$-$100} & 100  \\
Ljubljana  & {$-$\,45}           & 90   & {$-$5}  & 5   & {$-$5}  & 5   & {$-$25}  & 25   & 700  & 800  & {$-$25}  & 25   \\
Brigham    & {\phantom{$-$}\,55} & 305  & {$-$45} & 45  & {$-$15} & 15  & {$-$200} & 200  & 500  & 1000 & {$-$200} & 200  \\
\bottomrule
\end{tabular}
}
\caption{Parameter ranges used to train pose regression models for every dataset.}
\label{tab:parameters}
\vspace{-1em}
\end{table*}

\paragraph{Considerations.}
Manually selected fiducials chosen to evaluate mPD, mRPE, and mTRE metrics typically correspond to important anatomical landmarks. The geometry and local image contrast that make these fiducials easy for humans to identify in 2D and 3D imaging may also induce implicit bias towards these structures for registration error estimation at the expense of other anatomical regions. This makes $\mathcal L_{\mathrm{dGeo}}$ an important complementary error metric, as it does not rely on manually selected fiducials and instead assesses mis-registration across the full anatomical region.

\begin{figure*}[!t]
    \centering
    \includegraphics[width=\linewidth]{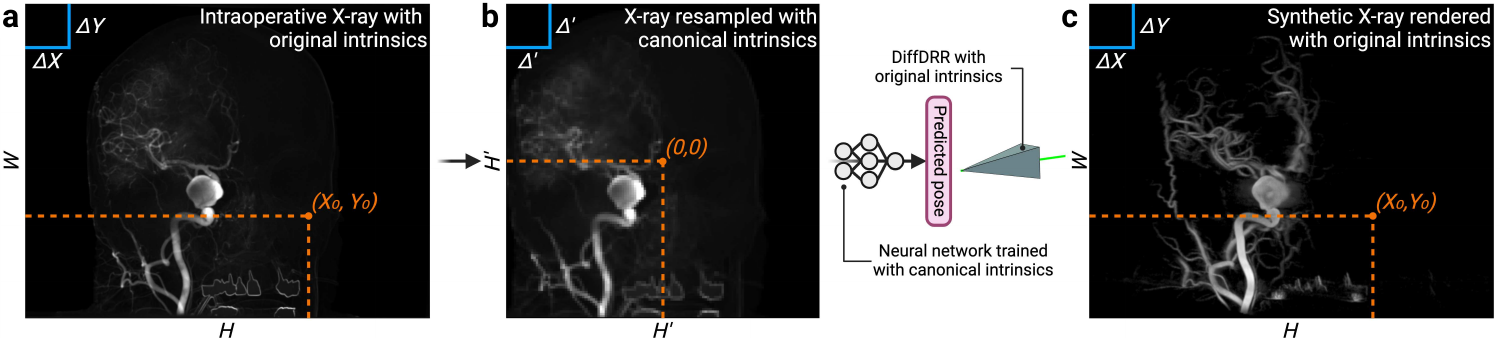}
    \caption{\textbf{Making pose regression neural networks insensitive to intrinsic parameter changes.} \textbf{a,} Using simple resampling operations (\ie translation, cropping, and bilinear interpolation), an intraoperative X-ray image with some set of intrinsic parameters---height $H$, width $W$, pixel spacing $(\Delta X, \Delta Y)$, principal point $(X_0, Y_0)$, and source-to-detector distance---is resampled to the canonical intrinsics used for rendering synthetic X-rays when training the patient-specific neural network. \textbf{b,} As a result, the resampled X-ray has a spatial resolution that matches the network's training data. \textbf{c,} After this network predicts the pose of the image, \name can render the predicted X-ray with the intrinsic parameters of the original image. That is, the synthetic X-ray is rendered with the original high-resolution for pose refinement via iterative optimization.}
    \label{fig:resample}
    \vspace{-1em}
\end{figure*}

\section{Training patient-specific neural networks}
\label[meth]{sec:simulation}

We train a patient-specific pose regression network using synthetic X-rays rendered from the patient's preoperative volume. To do this, we sample random C-arm poses from a distribution over plausible angles that may be acquired intraoperatively. Specifically, we sample individual pose parameters from the uniform distributions
\begin{align*}
    \alpha \sim \mathcal U[\alpha_{\mathrm{min}}, \alpha_{\mathrm{max}}] \quad&\quad x \sim\mathcal U[x_{\mathrm{min}}, x_{\mathrm{max}}] \\
    \beta \sim \mathcal U[\beta_{\mathrm{min}}, \beta_{\mathrm{max}}] \quad&\quad y \sim\mathcal U[y_{\mathrm{min}}, y_{\mathrm{max}}] \\
    \gamma \sim \mathcal U[\gamma_{\mathrm{min}}, \gamma_{\mathrm{max}}] \quad&\quad z \sim\mathcal U[z_{\mathrm{min}}, z_{\mathrm{max}}]
\end{align*}
and combine parameters into a single pose $\mathbf T$, where the rotation matrix $\mathbf R$ is given by \cref{eq:c-arm-rot} and the translation is defined as $\mathbf t = [x~y~z]^T$. The parameter ranges used in our experiments are provided in~\cref{tab:parameters}.

Given a batch of random C-arm poses $\mathbf T_n$, we generate a batch of synthetic images
\begin{equation}
    \mathbf I_n = \mathcal P(\mathbf T_n) [\mathbf V] \quad \forall \hspace{0.25em} n \in \{1, \dots, N\} \,,
\end{equation}
using our differentiable X-ray renderer~(\cref{fig:supp_renderer}).
If a 3D segmentation mask $\mathbf M$ assigning each voxel to a unique anatomical structure is provided, we also generate synthetic 2D masks:
\begin{equation}
    \mathbf S_n = \mathcal P(\mathbf T_n) [\mathbf M] \quad \forall \hspace{0.25em} n \in \{1, \dots, N\} \,.
\end{equation}
After generating these synthetic images, we execute the following training loop~(\cref{fig:wbct}f): (i) the batch of images is passed to a convolutional neural network $f_\theta : \mathcal I \mapsto \SE3$, which regresses a C-arm pose from each image $\hat{\mathbf T}_n = f_\theta(\mathbf I_n)$; (ii) these predicted poses are then passed back to our renderer to generate estimated X-rays $\hat{\mathbf I}_n = \mathcal P(\hat{\mathbf T}_n) [\mathbf V]$ and the optional 2D masks $\hat{\mathbf S}_n = \mathcal P(\hat{\mathbf T}_n) [\mathbf M]$; (iii) the estimated C-arm poses and X-rays are compared to the ground truth values produced via simulation to compute a loss function for the neural network:
\begin{equation}
\label{eq:loss}
\begin{split}
\mathcal{L}_n(\theta)
    &= \lambda_1\, \mathcal{L}_{\mathrm{dGeo}}(\mathbf{T}_n, \hat{\mathbf{T}}_n) \\
    &+ \lambda_2\, \mathcal{L}_{\mathrm{mNCC}}(\mathbf{I}_n, \hat{\mathbf{I}}_n)  \\
    &+ \lambda_3\, \mathcal{L}_{\mathrm{Dice}}(\mathbf{S}_n, \hat{\mathbf{S}}_n)\,,
\end{split}
\end{equation}
where $\mathcal L_{\mathrm{dGeo}}$ is the double geodesic distance between two poses in $\SE3$ in \cref{eq:dgeo}. $\mathcal L_{\mathrm{mNCC}}$ is the multiscale normalized cross correlation loss function~\cite{gopalakrishnan2024intraoperative}, and $\mathcal L_{\mathrm{Dice}}$ is a measure of the disagreement of two labelmaps~\cite{milletari2016v}. Across all experiments, we set $\lambda_1 = 0.01$, $\lambda_2 = 1$, and $\lambda_3 = 0.1$.

\paragraph{Data augmentation.}
In \cref{fig:supp_training}, we visualize a batch of synthetic X-rays rendered from the CT scan of a subject in the DeepFluoro dataset. During training, we heavily augment these images to simulate various intraoperative aberrations. We apply various intensity modifications, such as random local contrast equalization, blurring, sharpening, and additive Gaussian noise. Additionally, we implement random crops to simulate intraoperative changes to X-ray collimation or the presence of non-anatomical variations, such as surgical tools. Importantly, we do not apply any geometric image augmentations (\eg affine warps or vertical and horizontal flips) as this would alter the ground truth pose parameters that we aim to regress. As a result of these heavy augmentations, our pose regression models are insensitive to many domain shifts that occur in intraoperative imaging and produce consistently accurate initial pose estimates.

\paragraph{Network architecture.}
Let $\mathcal I$ represent the space of all synthetic and real X-ray images. We implement pose regression networks as $f_\theta = g \circ \mathcal E_\theta$, where $\mathcal E_\theta : \mathcal I \mapsto \mathbb R^{d+3}$ is a convolutional neural network (CNN) backbone and $g : \mathbb R^{d+3} \mapsto \SE3$ is a deterministic mapping from Euclidean space to the space of all C-arm poses. The Euclidean embedding produced by the CNN represents the rotational pose parameters ($\mathbb R^d$) and the translational pose parameters ($\mathbb R^3$). As an example, for the pose parameters used to generate random poses for pretraining, $g(\alpha, \beta, \gamma, x, y, z)$ is defined as
\begin{equation}
    \label{eq:random-pose}
    \begin{bmatrix}
        \mathbf R(\alpha, \beta, \gamma) & \mathbf R(\alpha, \beta, \gamma) \big( x\hat{\mathbf i} + y\hat{\mathbf j} + z\hat{\mathbf k} \big) \\
        \mathbf 0^T & 1
    \end{bmatrix} \,,
\end{equation}
where $\mathbf R(\alpha, \beta, \gamma)$ is defined in \cref{eq:c-arm-rot} and the form of this random pose is given by \cref{eq:camera-to-world}. $\mathcal E_\theta$ is implemented using a ResNet34 backbone~\cite{he2016deep}. All synthetic X-rays were rendered at $128 \times 128$ pixels using trilinear interpolation with a batch size of 116 X-rays per iteration. We train all patient-specific pose regression models on a single NVIDIA RTX 6000 Ada GPU (48 GB VRAM).

\begin{figure*}[t!]
    \centering
    \includegraphics[width=0.8\linewidth]{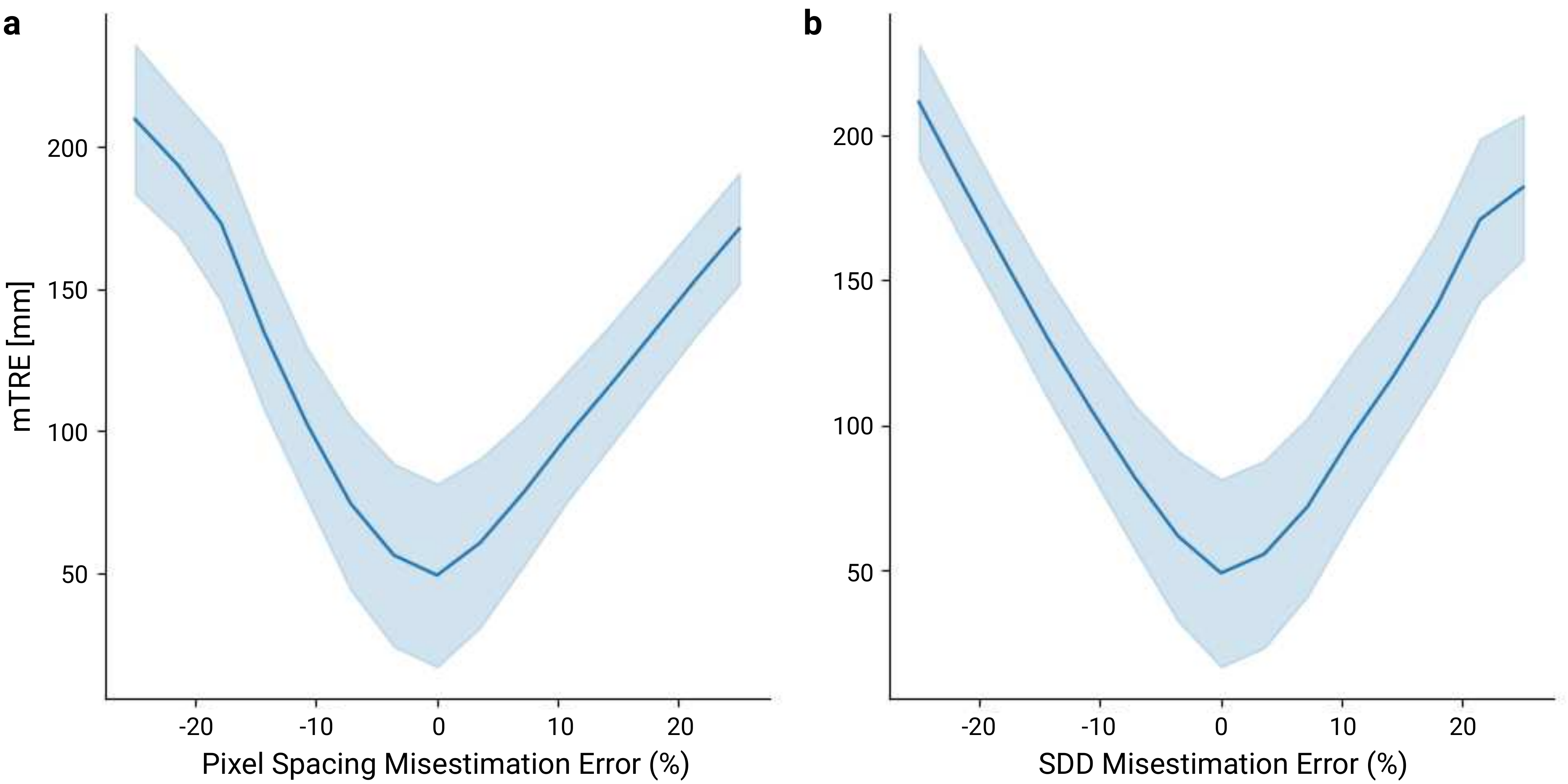}
    \caption{\textbf{Effect of misestimated intrinsics on initial pose estimation error.} Mean target registration error (mTRE) as a function of systematic error in C-arm intrinsic parameters for patient-specific models evaluated on the DeepFluoro dataset. \textbf{a,} Pixel spacing misestimation across \qty{0.194 +- 0.0485}{\mm} (\qty{\pm 25}{\percent} of nominal value). \textbf{b,} Source-to-detector distance (SDD) misestimation across \qty{1020 +- 255}{\mm} (\qty{\pm 25}{\percent} of nominal value). Shaded regions represent 95\% confidence intervals.}
    \label{fig:supp_intrinsics}
    \vspace{-1em}
\end{figure*}

\section{Adapting to variable intrinsic parameters}
\label[meth]{sec:resampling}
Neural networks used to regress landmark locations or C-arm poses from X-rays---including our patient-specific models---are typically trained using a fixed set of intrinsic parameters (\eg source-to-detector distance, detector height and width, pixel spacing, \etc). However, this is incongruous with clinical workflows, as the models cannot adapt to the interventionalist's changing image acquisition parameters on-the-fly (\eg panning the C-arm detector or narrowing the field of view to better visualize a particular structure). This clinical reality is not reflected in DeepFluoro, a cadaver study where the intrinsic parameters were identical for every image. However, since the DSA images in the Ljubljana dataset were acquired during real interventions, each image has different intrinsics.

To address this intraoperative challenge, we developed a simple geometric procedure for resampling an acquired X-ray image with a given set of intrinsic parameters to a canonical set of intrinsics using only basic image processing operations~(\cref{fig:resample}, \textit{left}). 
Specifically, we resample an input intraoperative image $\mathbf I(\mathbf x)$ as $\tilde{\mathbf{I}}(\tilde{\mathbf{x}})$ where
\begin{equation}
    \tilde{\mathbf{x}} = \tilde{\mathbf{K}} \mathbf{K}^{-1} \mathbf x \,.
\end{equation}
Here, $\mathbf{K}$ and $\tilde{\mathbf{K}}$ denote the intraoperative and canonical intrinsic matrices, respectively, and $\tilde{\mathbf{K}} \mathbf{K}^{-1}$ is a homography~\cite{hartley2003multiple}. This enables the neural network to perform pose regression independent of changing intrinsic parameters. While intraoperative images are resampled for initial pose estimation by the neural network, we render synthetic X-rays with the original intrinsic parameters during intraoperative pose refinement~(\cref{fig:resample}, \textit{right}). 

This approach is inherently different from training a neural network to be insensitive to changes in the image acquisition, for example, by rendering synthetic X-rays with varying intrinsic parameters. This strategy requires including five additional degrees of freedom in the parameter space during simulation. As we currently randomize the six pose parameters, including additionally simulating the intrinsic parameters would dramatically increase the training time. Furthermore, ground-truth intrinsic parameters are available from the Digital Imaging and Communications in Medicine (DICOM) header, allowing us to directly utilize known geometric information rather than training a model to recapitulate recorded measurements.

\section{Sensitivity to C-arm intrinsic parameters}
\label[meth]{sec:intrinsics}
2D/3D registration can be sensitive to the accuracy of C-arm intrinsic parameters. To quantify this sensitivity, we performed a simulation study in which we systematically perturbed either the pixel spacing or the source-to-detector distance (SDD) provided as input to the pose regression neural network. We evaluated our patient-specific models on the DeepFluoro dataset while varying both parameters across a range of \qty{\pm 25}{\percent} error. For pixel spacing, the tested range was \qty{0.194 +- 0.0485}{\mm}; for SDD, the tested range was \qty{1020 +- 255}{\mm}. As shown in~\cref{fig:supp_intrinsics}, misestimation of C-arm intrinsic parameters induces substantial errors in pose estimation. The model exhibits particularly high sensitivity to pixel spacing errors, where even a hundredth of a millimeter deviation increased the mTRE by tens of millimeters, with mTRE reaching approximately \qty{50}{\mm} at the optimal value and exceeding \qty{175}{\mm} at \qty{\pm 25}{\percent} error (\cref{fig:supp_intrinsics}a). In contrast, the model was more robust to absolute misestimation error in the SDD, producing comparably accurate pose estimates at \qty{50}{\mm} perturbations~(\cref{fig:supp_intrinsics}b). This finding is clinically relevant, as pixel spacing can typically be determined with high precision from the known specifications of the X-ray detector (sensor dimensions and pixel array size), whereas SDD may vary more substantially across different imaging configurations.

\begin{figure*}[hbt!]
    \centering
    \includegraphics[width=\linewidth]{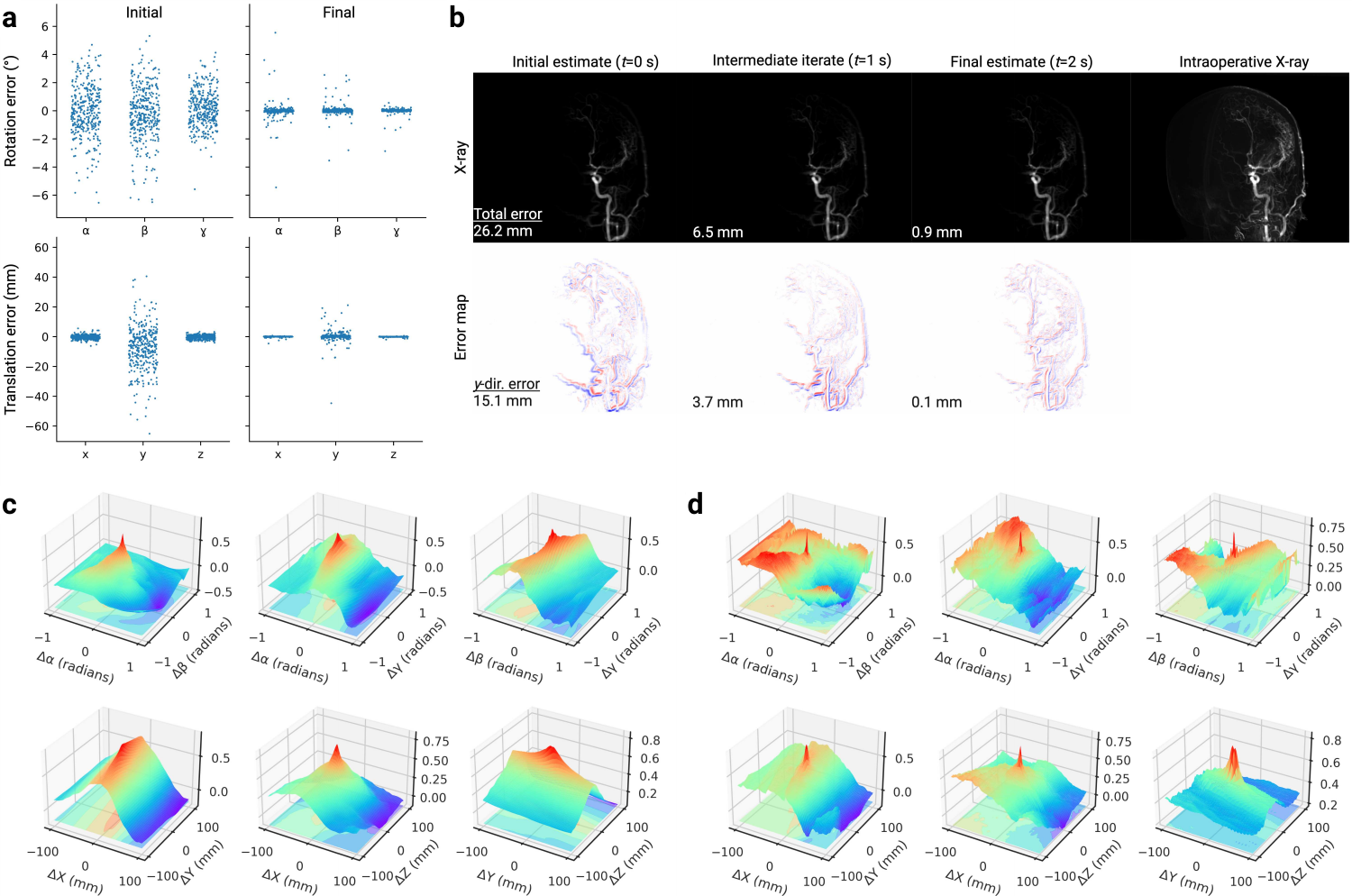}
    \caption{\textbf{Intraoperative pose refinement strategy.} \textbf{a,} Estimation errors of the six pose parameters before (\textit{left}) and after (\textit{right}) iterative optimization. All rotational parameters incur roughly \ang{\pm 2.5} of error and in-plane translational parameters ($x$ and $z$) incur roughly \qty{\pm 1.5}{\mm} of error. However, the source-to-object distance ($y$) incurs errors of \qty{\pm 15}{\mm}, demonstrating the difficulty in accurately estimating depth. \textbf{b,} Intraoperative pose refinement takes about \qty{2}{\second} and successfully overcomes the depth error in the network's initial pose estimate. \textbf{c,} The loss landscape induced by multiscale normalized cross correlation (mNCC) is smooth in a large neighborhood around the true pose, broadening the capture radius of our pose refinement strategy. However, mNCC is relatively non-specific about the true pose. \textbf{d,} In contrast, the loss landscape induced by gradient normalized cross correlation (gNCC) results in a more specific optimum at the expense of a rougher landscape further from the true pose. Averaging mNCC and gNCC achieves millimeter-accurate pose refinement.}
    \label{fig:ttopt}
    \vspace{-1em}
\end{figure*}

\section{Pose refinement}
\label[meth]{sec:optimization}
While neural networks in \name typically produce pose estimates within \qtyrange{20}{30}{\mm} of the ground truth pose, this error is not uniform across all the degrees of freedom that constitute the C-arm's pose. In \cref{fig:ttopt}a, \textit{left}, we visualize the distribution of errors for each degree of freedom in initial pose estimates (the geometric meaning of each parameter is illustrated in \cref{fig:overview}c). While the rotational degrees of freedom each have roughly \ang{\pm 2.5} of initial misestimation error (\cref{fig:ttopt}a, \textit{top left}), the translational degrees of freedom display much greater heterogeneity. For example, the in-plane translations ($x$ and $z$) incur errors of only \qty{\pm 1.5}{\mm}, but the source-to-object distance ($y$) incurs errors of \qty{\pm 15}{\mm} (\cref{fig:ttopt}a, \textit{bottom left}). This is because the resulting X-ray image changes very subtly for even large changes in the distance from the X-ray source to the object. To illustrate this phenomenon, we show an example of the iterative optimization process in \cref{fig:ttopt}b. While the different iterates do not appear particularly visually distinct, the registration error decreases from \qty{26.2}{\mm} to \qty{0.9}{\mm} after iterative optimization~(\cref{fig:ttopt}b, \textit{top}). In particular, the error of the estimated source-to-object distance reduces from \qty{15.1}{\mm} to \qty{0.1}{\mm}~(\cref{fig:ttopt}b, \textit{bottom}). Furthermore, this example highlights the inherent difficulty of attempting to perform manual 2D/3D registration, as all the synthetic X-rays appear visually similar to the intraoperative X-ray despite wildly different poses.

\paragraph{Image similarity metric.}
We tested multiple image similarity metrics to determine which enables the precise recovery of the true values of all pose parameters. The first metric we considered was multiscale normalized cross correlation (mNCC)~\cite{gopalakrishnan2024intraoperative}, which has previously been shown to increase the capture radius of 2D/3D registration via iterative optimization~\cite{gopalakrishnan2024intraoperative}. Second, we considered gradient normalized cross correlation (gradNCC)~\cite{grupp2018patch}, an image similarity metric that computes the correlation between Sobel-filtered versions of the two images. As this metric encourages the alignment of the edges in the two images, it is better suited for depth estimation than mNCC~\cite{grupp2020automatic}. 

To evaluate the behavior of these image similarity metrics, we visualize their loss landscapes~(\cref{fig:ttopt}c,d). Specifically, we render synthetic X-rays at perturbations from the ground truth pose (\ang{\pm60} for rotational parameters and \qty{100}{\mm} for translational parameters) and measure their similarity to the intraoperative X-ray. The loss landscape for mNCC is very smooth in this large region around the ground truth pose, which means that it is an ideal objective function to optimize when the initial pose estimate has high error~(\cref{fig:ttopt}c). However, this visualization also shows that the gradient of mNCC in the $y$-direction is relatively small. Whereas each of the other pose parameters has well-defined peaks, which lead to more precise and efficient optimization, the loss landscape of mNCC in the $y$-direction is relatively saddle-like, suggesting that pose refinement with mNCC may misestimate the source-to-object distance of an intraoperative X-ray by a few millimeters. In contrast, gradNCC has a much sharper landscape about the true source-to-object distance~(\cref{fig:ttopt}d). However, unlike mNCC, gradNCC is much less smooth far away from the ground truth pose and is therefore less robust to poor initial pose estimates. To combine the advantages of both metrics, we perform pose refinement by optimizing the average of mNCC and gradNCC, enabling robust and precise 2D/3D registration~(\cref{fig:benchmark}).

\paragraph{Structure-specific registration.}
If a segmentation map is available for the preoperative volume, we can use the ability of our renderer to generate X-ray images of specific parts of the preoperative volume to register individual anatomical structures. Specifically, for pelvic registration, we only render the left and right hips, sacrum, and L5 vertebra. By modeling those objects as a single rigid object, we make our registration framework insensitive to irrelevant domain shifts, such as the motion of the femur between the preoperative and intraoperative imaging. While segmentation labels are relatively easy to derive for most structures on CT and MRI thanks to open-source tools such as TotalSegmentator~\cite{wasserthal2023totalsegmentator} or MOOSE~\cite{ferrara2025sharing, sundar2022fully}, other structures like vessels remain difficult to segment. Therefore, for neurovascular registration, we do not use segmentation labels to guide registration. Despite this, we still achieve accurate results. 

\paragraph{Multiscale registration.}
To improve intraoperative registration speed, we implemented a multiscale rendering protocol that registers X-ray images at progressively higher resolutions. Specifically, when optimization of the image similarity metric plateaus at a particular scale, we progress to the next higher resolution. This coarse-to-fine registration strategy enables simultaneous optimization of global anatomical misalignment and local refinement of fine structures.

\begin{table*}[t]
\centering
\renewcommand{\arraystretch}{1.0}
\begin{tabular}{@{}c c c c c c c c@{}}
\toprule
 & Rank & Method & Renderer & mTRE [mm] & $z$ & $p_{\mathrm{adj}}$ & Significance \\
\midrule
 \multirow{7}{*}{\rotatebox{90}{\footnotesize\textbf{DeepFluoro}}}
 & \#1 & Finetuned & \name & 1.1(1.3) & \phantom{$-$0}---\phantom{.00} & \phantom{<0.}---\phantom{0} & --- \\
 & \#1 & \textit{De novo} & \name & 1.2(1.3) & \phantom{$-$}\phantom{0}0.08 & \phantom{<}1.0000 & n.s. \\
 & \#3 & Foundation & \name & 1.3(3.0) & \phantom{$-$}\phantom{0}7.34 & <0.0001 & *** \\
 & \#4 & Landmark init.~\cite{grupp2020automatic} & \textbf{xReg} & 3.8(4.1) & \phantom{$-$}16.29 & <0.0001 & *** \\
 & \#5 & Landmark reg.~\cite{grupp2020automatic} & \textbf{xReg} & 4.3(52.3) & \phantom{$-$}18.56 & <0.0001 & *** \\
 & \#6 & Fixed init.~\cite{grupp2019pose} & \textbf{xReg} & 365.9(329.5) & \phantom{$-$}42.87 & <0.0001 & *** \\
 & \#6 & Fixed init.~\cite{gopalakrishnan2022fast} & \name & 352.8(203.3) & \phantom{$-$}47.30 & <0.0001 & *** \\
\midrule
 \multirow{3}{*}{\rotatebox{90}{\footnotesize\textbf{Femur}}}
 & \#1 & Finetuned & \name & 8.0(5.7) & \phantom{$-$0}---\phantom{.00} & \phantom{<0.}---\phantom{0} & --- \\
 & \#1 & \textit{De novo} & \name & 8.0(5.8) & \phantom{$-$}\phantom{0}0.82 & \phantom{<}1.0000 & n.s. \\
 & \#3 & Foundation & \name & 343.1(647.4) & \phantom{$-$}15.09 & <0.0001 & *** \\
\midrule
 \multirow{3}{*}{\rotatebox{90}{\footnotesize\textbf{Ljubljana}}}
 & \#1 & Finetuned & \name & 1.4(1.8) & \phantom{$-$0}---\phantom{.00} & \phantom{<0.}---\phantom{0} & --- \\
 & \#1 & \textit{De novo} & \name & 1.3(1.5) & \phantom{0}$-$1.09 & \phantom{<}0.8262 & n.s. \\
 & \#3 & Foundation & \name & 1351.1(320.1) & \phantom{$-$}21.78 & <0.0001 & *** \\
\bottomrule
\end{tabular}
\caption{Ranking of each method's final pose estimation error per dataset, reported as median(IQR) in mm. $z$ and $p_{\mathrm{adj}}$ are from pairwise linear mixed-effects models comparing each method to Finetuned (reference) on log(mTRE) with subject as random intercept and X-ray as a variance component, accounting for the paired design where every X-ray is evaluated by every method. $p$-values are Bonferroni-corrected within each dataset. Ranks are derived from all pairwise comparisons; tied ranks indicate no significant difference ($p_{\mathrm{adj}} > 0.05$).}
\label{tab:ranks}
\end{table*}

\section{Transfer learning}
\label[meth]{sec:transfer}
Reusing the weights of a 2D/3D registration foundation model enables rapid patient-specific finetuning. However, this introduces an additional complexity: poses estimated by the foundation model are in the reference frame of the pretraining corpus, not the preoperative volume of interest. To account for this mismatch, we correct the pose estimates produced by the foundation model by mapping the patient-specific preoperative volume to the registration template. Specifically, we use Greedy~\cite{yushkevich2016ic} to rigidly align all volumes from the DeepFluoro, Ljubljana, Femur, and Brigham datasets to our whole-body template~(\cref{fig:wbct}e).

\section{Parsing partial pose parameters in DICOM headers}
\label[meth]{sec:dicom}
 Imaging data stored as DICOM files contain additional metadata beyond the raw pixel intensities of the image. In particular, certain X-ray DICOM formats encode the primary and secondary positioner angles of the C-arm, which correspond to the rotational parameters $\alpha$ and $\beta$, and the Source-to-Patient Distance attribute, which represents the $y$-direction translation~(\cref{fig:overview}c). These parameters comprise only three of the six degrees of freedom required to describe a C-arm pose as formulated in \cref{eq:random-pose}. Furthermore, while this partially specifies the pose of the C-arm, this formulation does not account for the position of the patient relative to the C-arm. That is, unless the isocenter of the patient's CT is perfectly aligned at the C-arm's center of rotation, the pose constructed from the DICOM metadata will contain misidentified translations.

\section{Model size and computational resources}
\label[meth]{sec:compute}
All our neural networks use a ResNet34 backbone (21,278,400 parameters) and two linear layers for rotation and translation regression (6,669 parameters). Our foundation model was trained on a single NVIDIA H200 GPU (141 GB of VRAM). All other models were trained on a single NVIDIA RTX 6000 Ada GPU (48 GB VRAM) in a 48‑core (96‑thread) Linux system with dual‑socket Intel(R) Xeon(R) Gold 5418Y CPUs (24 cores per socket, up to 3.80 GHz). With the exception of the Brigham dataset, pose refinement for all tested methods was performed on the same computational platform as used for network training (NVIDIA RTX 6000 Ada GPU). Due to privacy constraints, the Brigham dataset was registered on a secure workstation equipped with a single NVIDIA GeForce RTX 3060 GPU (12 GB VRAM).

\section{Statistical testing}
\label[meth]{sec:stats}
We now assess pairwise differences between registration methods for statistical comparison. The samples in our evaluation datasets are not independent as multiple X-rays are acquired from individual subjects to assess performance across poses, thereby necessitating modeling of repeat measurements. We therefore fit linear mixed-effects models on log-transformed mTRE results with the method as a fixed effect, the subject as a random intercept, and the X-ray as a variance component. The log transform was applied to stabilize variance and account for the right-skewed distribution of registration errors. The random intercept accounts for subject-level differences in difficulty, while the variance component for X-ray accounts for the paired design in which every X-ray is evaluated by every method, with some viewing angles being inherently harder to register than others. 

For each pair of methods, we fit a separate model and extract the Wald statistic ($z$) and two-sided p-value for the fixed effect of method, where the test statistic is assumed to follow a standard normal distribution asymptotically. $p$-values were Bonferroni-corrected within each dataset to control the family-wise error rate across all pairwise comparisons. Methods were ranked by their estimated marginal mean log(mTRE) from a single model fit to all methods simultaneously, with tied ranks assigned to methods whose pairwise difference did not reach significance after correction (adjusted $p > 0.05$). All models were fit using restricted maximum likelihood (REML) via the statsmodels Python package.

On DeepFluoro, our finetuned model achieves a median mTRE of \qty{1.1}{\mm} (IQR \qty{1.4}{\mm}). All baseline methods are statistically significantly worse (all Bonferroni-corrected $p < 0.0001$). There is no statistical difference between our finetuned and \textit{de novo} models (median \qty{1.2}{\mm}, IQR \qty{1.3}{\mm}; Bonferroni-corrected $p = 1.0$). The full pairwise analysis yields a clear ranking: patient-specific methods (rank \#1) > foundation model (rank \#3) > landmark initialization~\cite{grupp2020automatic} (rank \#4) > landmark regularization~\cite{grupp2020automatic} (rank \#5) > fixed initialization~\cite{grupp2019pose, gopalakrishnan2022fast} (rank \#6), with all differences significant after correction. On Femur and Ljubljana, our patient-specific models similarly achieve the lowest mTRE with statistical significance across all comparisons. While DeepFluoro contains only 6 subjects, the mixed-effects model explicitly accounts for this repeated-measures structure; the observed significance levels remain robust under the correctly modeled effective sample size.

\section*{Data availability}
We used the following public 2D/3D registration datasets: 
\begin{itemize}[noitemsep]
    \item DeepFluoro (\url{https://doi.org/10.7281/T1/IFSXNV})
    \item Ljubljana (\url{https://lit.fe.uni-lj.si/en/research/resources/3D-2D-GS-CA})
    \item Femur (\url{https://doi.org/10.1007/s11548-025-03391-4})
\end{itemize}
Remixed versions of these datasets in the DICOM format are available, with permission from the original authors, at \texttt{\href{https://huggingface.co/datasets/eigenvivek/xvr-data/tree/main}{https://huggingface.co/datasets/eigenvivek/xvr-data}}.
We used the following publicly available 3D imaging datasets:
\begin{itemize}[noitemsep]
    \item CTPelvic1K (\url{https://doi.org/10.5281/zenodo.4588402})
    \item NITRC MRA Atlas (\url{https://www.nitrc.org/projects/icbmmra})
    \item TotalSegmentator (\url{https://doi.org/10.5281/zenodo.6802613})
    \item FDG-PET-CT-Lesions (\url{https://doi.org/10.7937/gkr0-xv29})
\end{itemize}
Due to Health Insurance Portability and Accountability Act (HIPAA) regulatory requirements, the Brigham CTA/DSA and Boston Children's MRA/DSA datasets are unavailable for public release. 

\section*{Code availability}
The Python package and command line interface for \name, along with all scripts necessary to replicate the experiments presented in this manuscript, are available at \url{https://github.com/eigenvivek/xvr}. Pretrained foundation, \textit{de novo}, and finetuned model weights are available at \texttt{\href{https://huggingface.co/eigenvivek/xvr/}{https://huggingface.co/eigenvivek/xvr}}. \name is implemented in Python 3.10+ and uses PyTorch 2.2+~\cite{paszke2019pytorch} as its autodifferentiable backend.

\section*{Acknowledgments}
We are grateful to Theo van Walsum for explaining how C-arm poses are parameterized in DICOM headers. This work was supported by NIH NIBIB 5T32EB001680-19, the MIT CSAIL-Wistron Program, the MIT-IBM Watson AI Lab, the MIT Jameel Clinic, the MIT Health and Life Sciences Collaborative (HEALS), and the Chou Family Transformative Research Fund.

\section*{Contributions}
V.G. collected and standardized public data, developed code, trained models, ran experiments, analyzed results, created figures, and wrote the manuscript. All authors reviewed the manuscript and provided revisions and feedback. N.D., S.F., and P.G. provided technical advice. V.G. and N.D. found public benchmarking datasets. D.-D.G., A.A., A.M.L., and D.B.O. collected clinical data and provided clinical input and feedback. D.-D.G., A.M.L., and N.H. manually annotated clinical data. N.D. and P.G. served as co-principal investigators for this study and advised on technical details. P.G. provided research support for the project. No funders or third parties were involved in the study design, analysis, or writing.


\putbib
\end{bibunit}
\end{document}